\documentclass[superscriptaddress,nofootinbib,amsmath,amssymb,prc,showpacs,twocolumn]{revtex4-2}

\usepackage{graphicx}
\usepackage{dcolumn}
\usepackage{bm}
\usepackage{color}
\usepackage{ulem}

\usepackage[colorlinks]{hyperref}
\hypersetup{linkcolor=red,citecolor=blue,urlcolor=blue}

\usepackage{float,subfig,slashed,hyperref}

\begin{document}

\title{Quark-nuclear hybrid equation of state for neutron stars under modern observational constraints}

\author{G.A. Contrera}
\affiliation{CONICET, Godoy Cruz 2290, Buenos Aires, Argentina}
\affiliation{IFLP, UNLP, CONICET, Facultad de Ciencias Exactas, Diagonal 113 entre 63 y 64, La Plata 1900, Argentina}
\author{D. Blaschke}
\affiliation{Institute for Theoretical Physics, University of Wroclaw, Max Born Pl. 9, 50-204, Wroclaw, Poland}
\affiliation{Bogoliubov Laboratory of Theoretical Physics, Joint Institute for Nuclear research, Joliot-Curie str. 6, 141980 Dubna, Russia}
\affiliation{National Research Nuclear University (MEPhI), Kashirskoe Shosse 31, 115409 Moscow, Russia}
\author{J.P. Carlomagno}
\affiliation{CONICET, Godoy Cruz 2290, Buenos Aires, Argentina}
\affiliation{IFLP, UNLP, CONICET, Facultad de Ciencias Exactas, Diagonal 113 entre 63 y 64, La Plata 1900, Argentina}
\author{A.G. Grunfeld}
\affiliation{CONICET, Godoy Cruz 2290, Buenos Aires, Argentina}
\affiliation{Departamento de F\'\i sica, Comisi\'on Nacional de Energ\'{\i}a At\'omica, Av. Libertador 8250, (1429) Buenos Aires, Argentina}
\author{S. Liebing}
\affiliation{Bogoliubov Laboratory of Theoretical Physics, Joint Institute for Nuclear research, Joliot-Curie str. 6, 141980 Dubna, Russia}

\begin{abstract}
We study a family of equations of state for hybrid neutron star matter. 
The hybrid EOS are obtained by a Maxwell construction of the  first-order phase transition between a hadronic phase described by the relativistic density-functional EOS of the “DD2” class with excluded volume effects and a deconfined quark matter phase modeled by an instantaneous nonlocal version of the Nambu–Jona-Lasinio model in SU(2)$_f$ with vector interactions and color superconductivity.
The form factor in the nonlocal quark matter model is fitted to lattice QCD results in the Coulomb gauge.
Owing to strong coupling in the vector meson and diquark channels, a coexistence phase of color superconductivity and chiral symmetry breaking occurs.
Our results show an approximately constant behavior for the squared speed of sound with values of 0.4 - 0.6 in the density region relevant for neutron star interiors.
To simultaneously fulfill the constraints from the Neutron Star Interior Composition Explorer radius measurement for PSR J0740+6620 and tidal deformability from GW170817 it is necessary to consider a $\mu$-dependent bag pressure that mimics confinement. 
\end{abstract}

\maketitle

\section{Introduction\label{sect:intro}}
With the advent of multi-messenger observations of neutron stars (NSs) due to the detection of the gravitational wave signal from the inspiral phase of the binary NS merger GW170817 \cite{LIGOScientific:2017vwq,LIGOScientific:2017ync}, stringent new constraints for the equation of state (EOS) of NS matter appeared.
The measured low value of the tidal deformability $\Lambda=190^{+390}_{120}$
of NSs at a mass of 1.4\,$M_\odot$ that was inferred from GW170817 \cite{LIGOScientific:2018cki}
is a challenge for several stiff nuclear matter EOS such as DD2 \cite{Typel:2009sy}, a standard relativistic density functional (RDF) EOS. 
On the other hand, soft nuclear EOS like APR \cite{Akmal:1998cf} or SLy4 \cite{Chabanat:1997un}, which were favorably advocated in the GW170817 discovery paper \cite{LIGOScientific:2017ync},
fail in matching the recent constraint from the Neutron Star Interior Composition Explorer (NICER)-X-ray Multi-Mirror (XMM) mass-radius measurement on PSR J0740+6620 
that requires a large radius 
$R_{2.0}=13.7^{+2.6}_{-1.5}$\,km \cite{Miller:2021qha}
($R_{2.0}=12.39^{+1.30}_{-0.98}$\,km \cite{Riley:2021pdl}
from the Amsterdam team) at the mass 
$(2.08\pm0.07)\,M_\odot$ \cite{Fonseca:2021wxt}.
Besides this, these nuclear EOS have the caveat that they do not include the appearance of the strangeness degree of freedom due to the onset of hyperons that would occur at masses $\approx 1.5\,M_\odot$.

A successful description of the multi-messenger phenomenology of NSs including the recent constraints on $R_{1.4}$ and $R_{2.0}$ was given recently with hybrid stars based on the constant speed of sound (CSS) model for quark matter. 
A few examples can be found in Refs.  \cite{Drischler:2020fvz,Cierniak:2021knt,Somasundaram:2021ljr,Antic:2021zbn,Li:2021sxb}.
Most of them choose the value of the squared sound speed $c_s^2$ freely; others vary the onset of the deconfinement transition and $c_s^2$ so that a sufficiently high maximum mass is obtained.
Allowing values up to the causality limit $c_s^2=1$, it is possible to reach maximum masses up to $4 M_\odot$ \cite{Drischler:2020fvz,Cierniak:2021knt}. 
Such ambiguity is not satisfactory.
Moreover, a justification for using the CSS model of the quark matter EOS is highly desirable.

This is the main motivation for the present work.
Following up on initial indications that the EOS of a nonlocal chiral quark model with an instantaneous, separable interaction potential can be nicely described by a CSS model \cite{Zdunik:2012dj} and that also a covariant formulation of such a nonlocal Nambu–Jona-Lasinio (NJL)-type model is nicely fitted by a CSS model \cite{Antic:2021zbn}, we investigate this relationship between modern NS phenomenology and a microphysical approach formulated by a chiral quark model Lagrangian. 

In this work we present a hybrid EOS construction in order to describe the transition from nuclear to quark matter (QM). 
We start from a chiral QM model with instantaneous nonlocal interactions in the scalar-pseudoscalar meson channel which predicts a three-momentum dependence of the quark mass compatible with Coulomb gauge lattice QCD (LQCD).
In addition, the model has a scalar diquark interaction channel that results in color superconductivity (2SC phase) and a repulsive vector interaction to provide sufficient stiffness of the QM at high densities in order to satisfactorily fulfill the observational constraint of a high neutron star mass exceeding $2~M_\odot$.
As it will be shown, this model produces an approximately constant speed of sound at the energy densities relevant for NS interiors.

Similar approaches for astrophysical applications of hybrid EOS within the (local) NJL model have been presented in the literature, considering either diquark interactions~\cite{Ippolito:2007hn,Baldo:2002ju,Buballa:2003et,Klahn:2013kga}, vector interactions~\cite{CamaraPereira:2016chj,Lourenco:2012yv,Hell:2014xva}, or both interaction currents ~\cite{Baym:2017whm,Baym:2019iky,Fujimoto:2019sxg}.
The ``instantaneous'' nonlocal versions of the NJL model including 2SC are presented, for instance, in Refs.~\cite{Grigorian:2003vi,Grigorian:2006qe}, and in Ref.~\cite{Klahn:2006iw} authors added vector interactions.
However, for nonlocal versions of the NJL model with a covariant form factor, hybrid EOS considering vector interactions were studied in Refs.~\cite{Orsaria:2012je,Orsaria:2013hna}, whereas the presence of diquarks and vector interactions were considered in Refs.~\cite{Blaschke:2007ri,Alvarez-Castillo:2018pve}.
By comparing the $M$-$R$ relations obtained in these works, the effect of including the superconducting phase is clearly seen: increasing the diquark coupling lowers onset mass for quark deconfinement. 

In Sec. II we describe the QM model and in Sec. III the construction of the hybrid EOS with its application to the modern neutron star phenomenological constraints on the mass-radius diagram and the tidal deformability.
Then in Sec. IV we present the summary and conclusions.
We added an Appendix with complementary calculations for the QM model, the Tolman-Oppenheimer-Volkoff (TOV) and tidal deformability equations.

\section{Instantaneous nonlocal chiral quark model with vector interactions and 2SC\label{sect:3DFFqm}}
We describe QM within a nonlocal chiral quark model which includes scalar and vector quark-antiquark interactions and antitriplet scalar diquark interactions.
The corresponding effective Euclidean action in the case of two light flavors is given by
\begin{eqnarray}
S_E &=& \int d^4 x \ \left\{ \bar \psi (x) \left(- i \rlap/\partial + m_c
\right) \psi (x) - \frac{G_S}{2} j^f_S(x) j^f_S(x) \right.
\nonumber \\
&-& \left. \frac{G_D}{2} \left[j^a_D(x)\right]^\dagger j^a_D(x) 
{+} \frac{G_V}{2} j_V^{\mu}(x)\, j_V^{{\mu}}(x) \right\} 
\label{action}
\end{eqnarray}
Here $m_c$ is the current quark mass, which is assumed to be equal for $u$ and $d$ quarks, whereas the currents $j_{S,D}(x)$ are given by nonlocal operators based on a separable approximation to the effective one gluon exchange (OGE) model of QCD~\cite{Blaschke:2007ri, GomezDumm:2005hy}.
In this work we use natural units with $\hslash = c = k_B = 1$.
The currents read
\begin{eqnarray}
j^f_S (x) &=& \int d^4 z \  g(z) \ \bar \psi(x+\frac{z}{2}) \ \Gamma_f\,
\psi(x-\frac{z}{2})\,,
\nonumber 
\\
j^a_D (x) &=&  \int d^4 z \ g(z)\ \bar \psi_C(x+\frac{z}{2}) \ i
\gamma_5 \tau_2 \lambda_a \ \psi(x-\frac{z}{2})\,, 
\nonumber
\\
{j^\mu_V (x)} &=& {\bar \psi(x)~ i\gamma^\mu\ \psi(x)}\,,
\label{cuOGE}
\end{eqnarray}
where we defined $\psi_C(x) = \gamma_2\gamma_4 \,\bar \psi^T(x)$ and $\Gamma_f=(\openone,i\gamma_5\vec\tau)$, while $\vec \tau$ and $\lambda_a$, with $a=2,5,7$, stand for Pauli and Gell-Mann matrices acting on flavor and color spaces, respectively.
The functions $g(z)$ in Eqs.~(\ref{cuOGE}) are nonlocal ``instantaneous'' form factors (3D-FF) characterizing the effective quark interaction, which depends on the spatial components of the momentum ($\vec{p}$).
Note that the vector current in Eq.~(\ref{cuOGE}) is considered to be local.
The reason will be given below.
The effective action in Eq.~(\ref{action}) might arise via Fierz rearrangement from some underlying more fundamental interactions, and is understood to be used $-$at the mean field level$-$ in the Hartree approximation. 
The vector mean field (MF) plays a special role since it affects the chemical potential and thus the baryon density which is a constraint and not a dynamic degree of freedom.
In general, the ratios of coupling constants $\eta_D = G_D/G_S$, $\eta_V =G_V/G_S$ would be determined by these microscopic couplings; for example, OGE interactions in the vacuum lead to $\eta_D =0.75$ and $\eta_V= 0.5$.
However, since the precise derivation of effective couplings from QCD is not known, there is a large theoretical uncertainty in these ratios.
Details of the values used in the present work will be given below.

We proceed by considering a bosonized version of this quark model, in which scalar, vector, and diquark fields are introduced.
Moreover, we expand these fields around their respective MF values, keeping the
lowest-order contribution to the thermodynamic quantities.
The only nonvanishing mean field values in the scalar and vector sectors correspond to isospin zero fields, $\bar\sigma$ and $\bar\omega$, respectively, while in the diquark sector, owing to the color symmetry, one can rotate in color space to fix $\bar\Delta_5=\bar\Delta_7=0$, $\bar\Delta_2=\bar\Delta$.

Now we consider the Euclidean action at both finite temperature $T$ and baryon
chemical potential $\mu_B$.
The simplicity of the 3D-FF is that the Matsubara summation can be performed analytically.
We introduce different chemical potentials $\mu_{fc}$ for each flavor and color.
In principle one has six different quark chemical potentials, corresponding to quark flavors $u$ and $d$ and quark colors $r,g$ and $b$.
However, there is a residual color symmetry (say, between red and green) arising from the direction of $\bar\Delta$ in color space.
Moreover, if we require the system to be in chemical equilibrium, it can
be seen that chemical potentials are not independent from each other.
In general, it is shown that all $\mu_{fc}$ can be written in terms of three independent quantities: the baryonic chemical potential $\mu_B$, a quark electric chemical potential $\mu_{Q_q}$ and a color chemical potential $\mu_8$.
The corresponding relations read
\begin{eqnarray}
\mu_{ur} &=& \mu_{ug} = \frac{\mu_B}{3} + \frac23 \mu_{Q_q} + \frac13
\mu_8 \nonumber \\ 
\mu_{dr} &=& \mu_{dg} = \frac{\mu_B}{3} - \frac13 \mu_{Q_q} + \frac13
\mu_8 \nonumber \\
\mu_{ub} &=& \frac{\mu_B}{3} + \frac23 \mu_{Q_q} -
\frac23 \mu_8 \nonumber \\
\mu_{db} &=& \frac{\mu_B}{3} - \frac13 \mu_{Q_q} -
\frac23 \mu_8 .
\label{chemical}
\end{eqnarray}
As we considered here the vector meson mean field $\bar{\omega}$, that comes from the term with $\gamma_0$ in the vector current in Eqs. (\ref{cuOGE}), the chemical potentials are shifted as
\begin{equation}
\tilde{\mu}_{fc} = \mu_{fc} - \bar{\omega}.
\end{equation}
Since we considered in Eqs.~(\ref{cuOGE}) that the vector current is local, the above shown renormalized chemical potentials only depend on the mean field $\bar{\omega}$. If we considered a nonlocal vector current, the term driven by $\bar{\omega}$ would have included the form factor, giving as a result, a chemical potential dependence on the three-momentum.

Following Ref.~\cite{Blaschke:2003yn}, it is convenient to define
\begin{equation}
\tilde{\mu}_c = \frac{\tilde{\mu}_{uc} + \tilde{\mu}_{dc}}{2}  
\end{equation}
and
\begin{equation}
\delta\tilde{\mu}_c = \frac{\tilde{\mu}_{uc} - \tilde{\mu}_{dc}}{2}.  
\end{equation}

Thus, the corresponding mean field grand canonical thermodynamic potential per unit volume can be written as 
\begin{eqnarray}
\Omega^{MFA}  =  &-& \frac{T}{V} \, 
\ln {\cal Z}^{MFA}  = \frac{ \bar
\sigma^2 }{2 G_S} + \frac{ {\bar \Delta}^2}{2 G_D} - \frac{\bar
\omega^2}{2 G_V}\ \nonumber \\ &-& 2 \int \frac{d^3 \vec{p}}{(2\pi)^3} \; 
\xi(\vec{p}) , 
\label{mfaqmtp}
\end{eqnarray}
where 
\begin{eqnarray}
\xi(\vec{p})  & &=  \sum_{\kappa,s=\pm} 2 \; \left\{ \epsilon _r^\kappa/2 \; + \; T \; \ln \left[ 
1 \; + \; e^{- \tfrac{\epsilon _r^\kappa + s\ \delta\tilde\mu_r}{T}}\right] \right\} \nonumber \\
&+& \sum_{\kappa,s=\pm} \; \left\{ \bar{E}_b^\kappa/2 \; + \; T \; \ln \left[ 
1 \; + \; e^{- \tfrac{\bar{E}_b^\kappa + s\ \delta \tilde\mu_b}{T}}\right] \right\}.
\label{integrando}
\end{eqnarray}

In the above expression the $\pm$ means that one has to consider two terms with each sign.
We defined
\begin{equation}
\epsilon _r^{\pm} = \bar{E}_r^{\pm} \sqrt{1 + {\left[{g(\vec{p}) \bar \Delta}/{\bar{E}_r^{\pm}}\right]^2}}
\end{equation}
for the gapped colors, where 
\begin{equation}
\bar{E}_r^{\pm} = E \pm \tilde{\mu}_r
\end{equation}
and
\begin{equation}
\bar{E}_b^{\pm} = E \pm \tilde{\mu}_b
\end{equation}
and for the ungapped ``blue'' color of quarks, and the dispersion relation is given by
\begin{equation}
E^{2} = \vec{p}~^2 + M^2(\vec{p}).
\end{equation}
Here, the momentum-dependent quark mass function is
\begin{equation}
M(\vec{p}) = m_c + g(\vec{p}) \bar\sigma.
\end{equation}

The mean field values $\bar \sigma$, $\bar \Delta$ are obtained from the coupled gap equations together with the constraint equation for $\bar \omega$:
\begin{eqnarray}
\frac{ \partial \Omega^{MFA}}{\partial \bar \sigma} = 0 \ , \ \ \
\frac{\partial \Omega^{MFA}}{\partial \bar \Delta} = 0 \ , \ \ \
\frac{\partial \Omega^{MFA}}{\partial \bar \omega} = 0 \ .
\label{gapeq}
\end{eqnarray}

Then the corresponding EOS can be obtained by inserting  into Eq.~(\ref{mfaqmtp}) the mean field values $\bar\sigma$, $\bar\omega$, and $\bar\Delta$ which are obtained by solving the gap equations of Eqs.~(\ref{gapeq}), explicitly shown in  Eqs.~(\ref{gap_sig})-(\ref{gap_ovec}), and using the regularization prescription of Eq.~(\ref{eq:omreg}).
Thus, the pressure of the quark matter is given by
\begin{equation}
    P_q = - \Omega^{MFA}_{reg} \,.
\end{equation}

Now, if we want to describe the behavior of quark matter in the core of neutron stars, in addition to quark matter we have to take into account the presence of electrons and muons.
Thus, treating leptons as a free relativistic Fermi gas, the total pressure of the quark matter plus leptons is given by
\begin{equation}
P = P_q + P_{lep} \ ,
\label{potential}
\end{equation}
where $P_{lep}$ reads
\begin{equation}
P_{lep} = 2  \; T\; \sum_{\stackrel{l = e, \mu}{s =\pm}}
\int \frac{d^3\vec{p}}{(2 \pi)^3} \ln \left[ 
1 \; + \; e^{- \tfrac{\epsilon_l + s \; \mu_e}{T}}\right]\,, 
\label{free}
\end{equation}
with $\epsilon_l = \sqrt{\vec{p}^2 + m_l^2} $ and the chemical  potential $\mu_e = \mu_{\mu}$.

In addition, it is necessary to take into account that quark matter has to be in $\beta$ equilibrium with electrons and muons through the $\beta$-decay reactions
\begin{equation}
d\to u+l+\bar\nu_l\ ,~~~  u+l\to d+\nu_l\ ,
\end{equation}
for $l=e,\mu$.
Thus, assuming that (anti)neutrinos escape from the stellar core, we have an additional relation between fermion chemical potentials, namely,
\begin{equation}
\label{betaeq}
\mu_{dc} - \mu_{uc} = - \mu_{Q_q} = \mu_l
\end{equation}
for $c=r,g,b$, $\mu_l = \mu_e = \mu_\mu $.

Finally, in the core of neutron stars we also require the system to be electric and color charge neutral; hence the number of independent chemical potentials reduces further.
Indeed, $\mu_l$ and $\mu_8$ get fixed by the condition that charge and color number densities vanish,
\begin{eqnarray}
n_{Q_{tot}} &=& n_{Q_q}- \sum_{l=e,\mu}n_l \nonumber \\
& =& \sum_{c=r,g,b} \left(\frac23 \ n_{uc} - \frac13 \ n_{dc} \right)
- \sum_{l=e,\mu}n_l \ = \ 0 \ \ , \nonumber \\
n_8 & = & \frac{1}{\sqrt3} \sum_{f=u,d}
\left(n_{fr}+n_{fg}-2n_{fb} \right) \ = \ 0 \ ,
\label{dens}
\end{eqnarray}
where the expressions for the different number densities can be found in Appendix \ref{sect:details_QM}. 

In summary, in the case of neutron star quark matter, for each value of $T$ and $\mu_B$ one can find the values of $\bar \Delta$, $\bar \sigma$, $\bar \omega$, $\mu_l$, and $\mu_8$ by solving Eqs.~(\ref{gapeq}), supplemented by Eqs.~(\ref{betaeq}) and~(\ref{dens}).
This allows to obtain the quark matter EOS in the thermodynamic region we are interested in.

The energy density can be written as
\begin{equation}
\varepsilon = -P + T\;s + G
\end{equation}
where $s = -\partial \Omega/\partial T$.
Here, $G$ is the Gibbs free energy, which depends on conserved charges:
\begin{eqnarray}
G &=& \sum_{\alpha} \mu_{\alpha} \; n_{\alpha}  \nonumber \\  
&=& \mu_B \; n_B + \mu_Q \; n_Q + \mu_8 \; n_8. 
\label{eq:Gibbs_energy}
\end{eqnarray}
By imposing electric charge and color charge neutrality, the last two terms of the above equation are zero; then $G$ can be written as
\begin{equation}
G =\sum_{f,c} \mu_{f,c}\;n _{f,c}  + \sum_{l = e,\mu} \mu_{l}\;n _{l},
\end{equation}
where $n_B = (1/3)(n_u + n_d)$ with $n_f = \sum_c n_{f,c}$ and the chemical potentials $\mu_{f,c}$ are defined in Eqs.~(\ref{chemical}).

\subsection{Zero temperature limit}
In the present work we are interested in describing hybrid EOS for cold compact stellar systems. Then, we will take the zero limit for the temperature.
The corresponding MF grand canonical thermodynamic potential per unit volume can be written as in Ref.~\cite{Blaschke:2003yn}
\begin{eqnarray}
\Omega^{MFA}_{T=0,reg}  &=&  \frac{ \bar \sigma^2 }{2 G_S} + \frac{ {\bar \Delta}^2}{2 G_D} - \frac{\bar
\omega^2}{2 G_V} \nonumber \\ 
&-&  \int \frac{d^3 \vec{p}}{(2\pi)^3} \xi_{T=0}(\vec{p})  \; + \Omega_{T=0}^{lep}
\label{mfaqmtp_0}
\end{eqnarray}
where,
\begin{eqnarray}
\xi_{T=0}(\vec{p}) &=& \sum_{\kappa,s=\pm} \left\{ 2 \left[|\epsilon^{\kappa}_r +s\ \delta\tilde\mu_r| - E_0 \right]  \right.
\nonumber \\ &+&   \left.  \left[|E^{\kappa}_b + s\ \delta\tilde\mu_b| - E_0 \right]  \right\}.  
\end{eqnarray}

For leptons, the thermodynamic potential at vanishing temperature is given by~\cite{Blaschke:2007ri}

\begin{equation}
\Omega^{lep}_{T=0} = - \frac{1}{24\pi^2} \sum_{l = e, \mu} \mu_l^4 F(m_l/\mu_l) ,
\label{freepot}
\end{equation}
with
\begin{equation}
F(x) = 5\left(\frac{2}{5}-x^2\right) \sqrt{1-x^2} + 3x^4 \ln \left[\left(1+\sqrt{1-x^2} \right)x^{-1} \right] \nonumber  
\label{freeomegareg}
\end{equation}

valid when $\mu_l > m_l$.

It should be noticed that, in general, there might be regions for which there is more than one solution for each value of $\mu_B$. 
The stable solution corresponds to an overall minimum of the thermodynamic potential, from which the expressions for some other relevant quantities can be easily derived.
The quark and lepton densities are defined as 
\begin{eqnarray}
n_{fc}  =  - \frac{\partial \Omega^{MFA}_{T=0,reg}}{\partial \mu_{fc}}\, , ~
n_l = - \frac{\partial \Omega^{MFA}_{T=0,reg}}{\partial \mu_{e}}.
\label{densities}
\end{eqnarray}
The quark chiral condensate is defined as
\begin{equation}
\langle \bar \psi \psi \rangle  =  \frac{ \partial \Omega^{MFA}_{T=0,reg}}{\partial m_c}.
\end{equation}
Finally, a magnitude which is important to determine the characteristic of the chiral phase transition is the chiral susceptibility $\chi$, and it can be calculated as
\begin{equation}
\chi = - \frac{ \partial^2 \Omega^{MFA}_{T=0,reg}} {\partial m_c^2} =
- \frac{ \partial \langle \bar \psi \psi \rangle} {\partial m_c}\ .
\end{equation}

\subsection{Form factors and set of parameters}
\label{setLQCD}
To fully specify the nonlocal NJL model under consideration, one has to fix the model parameters as well as the instantaneous form factor $g(\vec{p})$ that characterize the nonlocal interactions between quarks in both channels $q\bar{q}$ and $qq$.
In this work we consider an exponential momentum dependence for the form factor in momentum space,
\begin{eqnarray}
\label{ff}
g(\vec{p}) &=& \exp[-\vec{p}\,^2/\Lambda_0^2]\ . \nonumber 
\end{eqnarray}
This form, which is widely used, guarantees a fast ultraviolet convergence of quark loop integrals. 
Notice that the energy scale $\Lambda_0$ has to be taken as an additional parameter of the model.
Other functional forms, e.g., Lorentzian~\cite{GomezDumm:2006vz, GomezDumm:2010cta, Carlomagno:2018tyk} or Woods-Saxon ~\cite{Grigorian:2006qe} form factors, have also been considered in the literature, concluding that the form factor choice does not have a considerable impact on the qualitative predictions for the relevant thermodynamic quantities~\cite{Blaschke:2007ri, GomezDumm:2005hy, Blaschke:2003yn, Grigorian:2003vi, Carlomagno:2013ona}.

Given  the  form  factor  functions,  it  is  possible to  set the model parameters to reproduce the observed meson phenomenology.  

\begin{figure}[h]
    \centering
    \includegraphics[width=0.49\textwidth]{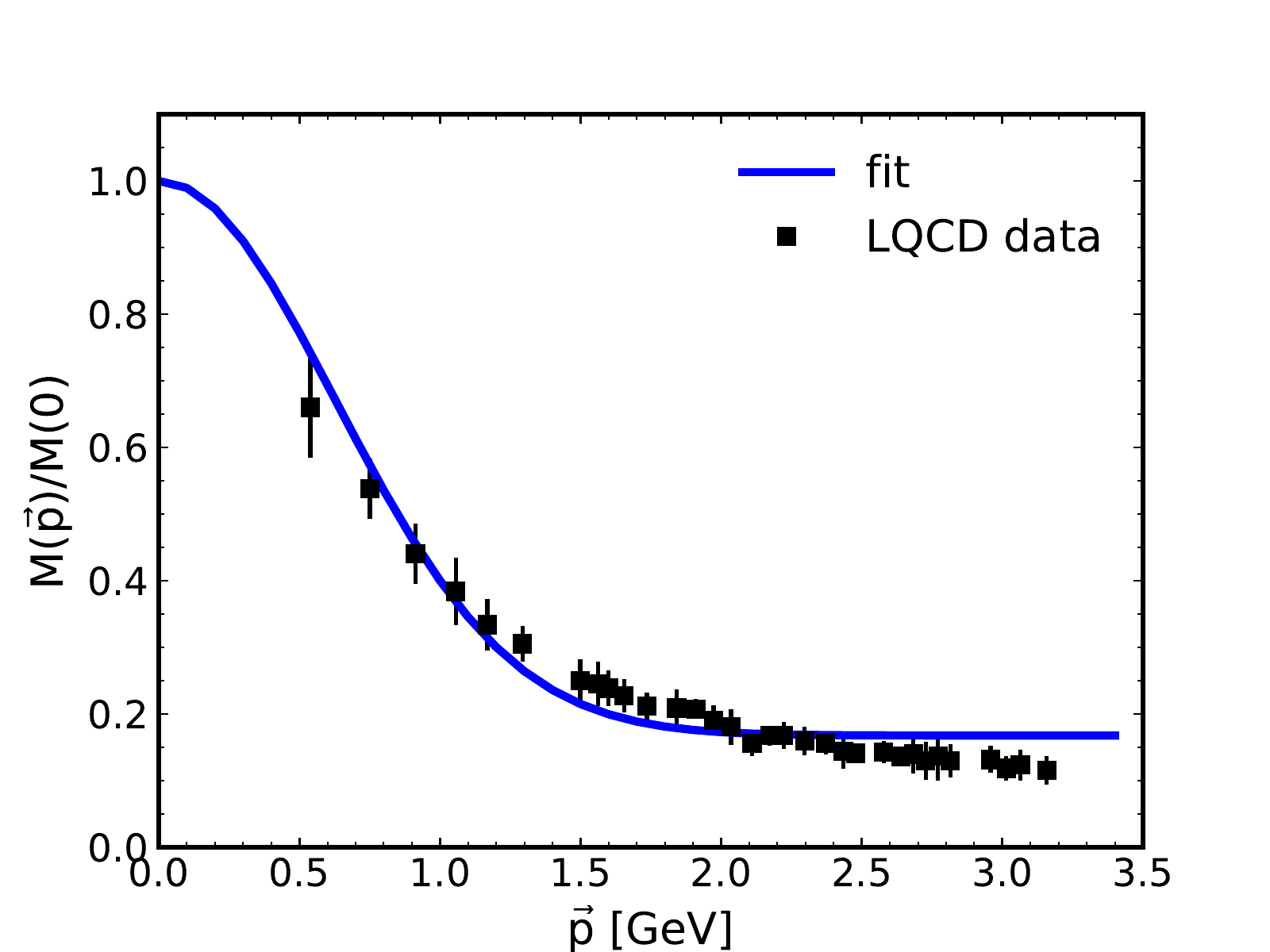}
    \caption{
    LQCD data from Ref.~\cite{Burgio:2012ph} and our fit for the normalized quark effective mass.}
    \label{fig:fits}
\end{figure}

First, we perform a fit to LQCD results (in the Coulomb gauge) from Ref.~\cite{Burgio:2012ph} for the normalized quark effective mass $M(\vec{p})$/$M(0)$.
This fit, quoted in Fig.~\ref{fig:fits}, has been carried out considering results up to 3 GeV, obtaining $\Lambda_0=885.47$~MeV.
\footnote{The normalization for the LQCD data was done using data from Fig.~7(d) of Ref.~\cite{Burgio:2012ph} together with Eq.~(26) evaluated at $k=0$ with the corresponding value of the LQCD bare quark mass.}

Finally, by  requiring  that  the  model  reproduce  the  empirical values of two physical quantities, chosen to be the pion mass $m_\pi=138$~MeV and the pion weak decay constant $f_\pi=92.4$~MeV, one can determine the remaining model parameters $m_c = 2.29$~MeV and $G_S=9.92$~GeV$^{-2}$.

\subsection{Phase diagram and speed of sound}
\begin{figure}[H]
    \centering
    \includegraphics[width=0.50\textwidth]{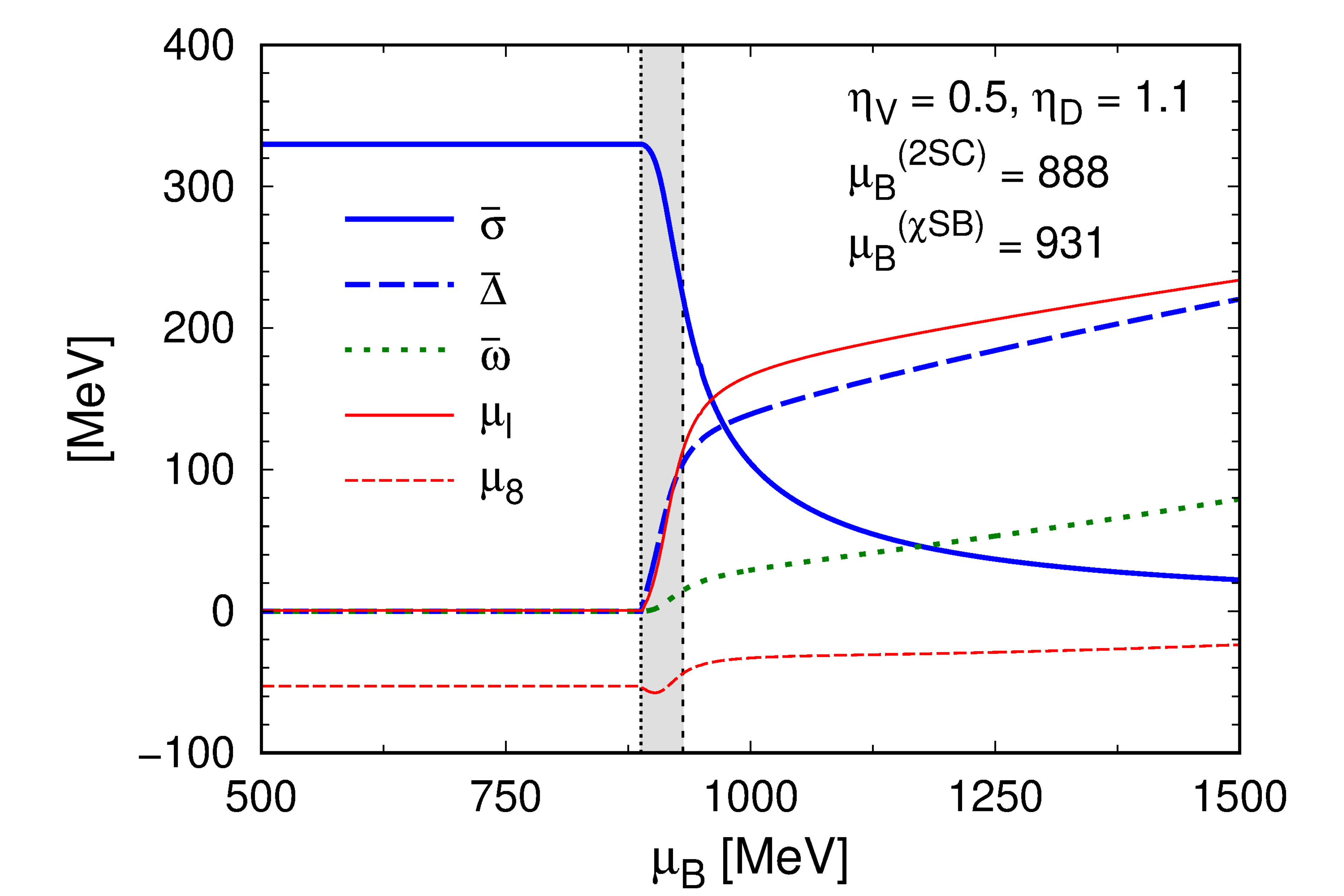}
    \caption{
   $\bar{\sigma}$, $\bar{\Delta}$, $\bar{\omega}$, $\mu_{l}$, and $\mu_8$ (in MeV) as functions of $\mu_B$ (in MeV).}
    \label{fig:VEVs}
\end{figure}

Let us start by studying the behavior of the mean field values for representative values of $\eta_D$ and $\eta_V$ as a function of the baryonic chemical potential.
Our results are shown in Fig.~\ref{fig:VEVs}.
There, as functions of $\mu_B$, we quote the MF values $\bar{\sigma}$, $\bar{\Delta}$, and $\bar{\omega}$ and the lepton and color chemical potentials, $\mu_{l}$ and $\mu_8$, for $\eta_D=1.1$ and $\eta_V=0.5$.
The critical chemical potentials $\mu_B^c$ are denoted with thin black vertical lines.
The dotted one, at $\mu_B=888$~MeV, indicates the baryonic chemical potential at which the total pressure vanishes.
Here, the mean field value of the diquark field vanishes, denoting a second-order phase transition.
However, at $\mu_B=931$~MeV, one finds the peak of the chiral susceptibility, indicating a crossover phase transition to a region where the chiral symmetry is partially restored.
In this narrow region of about 40 MeV, denoted by the grey band in the figure, one has a 2SC phase with a finite and small value of the diquark gap coexisting with the chiral symmetry breaking ($\chi$SB) phase. 

The appearance of a coexistence phase of 2SC and $\chi$SB has been discussed in the three-flavor case due to the mixing between $ud$-diquark condensate and the chiral condensate in the strange quark sector by the Fierz-transformed 't Hooft determinant interaction (axial anomaly; see \cite{Hatsuda:2006ps,Abuki:2010jq}).
However, in the two-flavor case considered here, the crossover occurs due to strong coupling and is related to the phenomenon of BEC-BCS crossover in strongly coupled fermion systems \cite{Zablocki:2009ds}.

Carrying out the previous analysis in the ranges allowed by the model for $\eta_D$ and $\eta_V$, namely, $0.0<\eta_V<1.2$ and $0.9<\eta_D<1.2$, we can set up the corresponding phase diagram and analyze the features of the phase transitions in the $\eta_D$-$\mu_B^c$ plane for different fixed values of $\eta_V$.

\begin{widetext}

\begin{figure}[!h]
    \centering
    \includegraphics[width=0.96\textwidth]{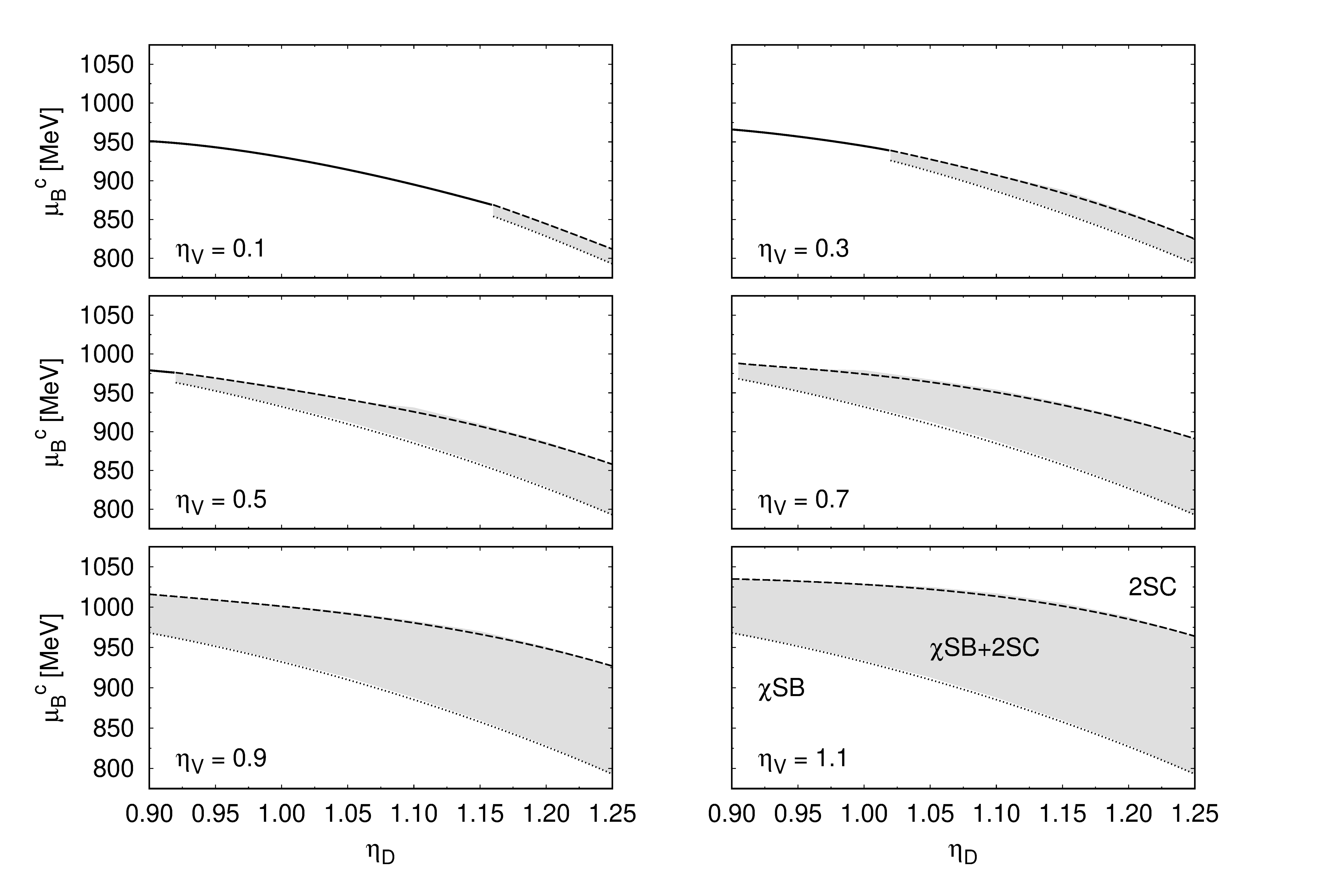}
    \caption{ \label{fig:phasediagram}  QM phase diagram in the $\eta_D$-$\mu_B^c$ plane. Solid, dotted, and dashed lines correspond to first, second, and crossover phase transitions. The shaded band indicates the coexistence region.}
\end{figure}

\end{widetext}

The phase diagram can be sketched by analyzing the numerical results obtained for the relevant order parameters. 
For the chiral symmetry restoration we take as order parameter the chiral quark condensate, while for the onset of the diquark condensation we take the MF value of the diquark field.
The chiral critical chemical potentials are defined by the positions of the peaks in the chiral susceptibilities in the region where the transition occurs as a smooth crossover, denoted by dashed lines in Fig.~\ref{fig:phasediagram}.
To sufficiently low diquark couplings the chiral restoration takes place as a first-order phase transition (solid lines in the figure).
However, when $\eta_D$  is increased, the chiral critical chemical potential gets reduced, and the chiral transition continues to be of first order up to a certain critical end point (CEP).
For larger values, the chiral restoration phase transition proceeds as a smooth crossover.

However, the diquark condensation is always a second-order phase transition, denoted by dotted lines in Fig.~\ref{fig:phasediagram}.

The region between both phase transition curves, denoted as a grey band in the figure, is a coexistence region where the chiral symmetry remains broken with a nonvanishing diquark MF value, the same as in  Fig.~\ref{fig:VEVs}.

Finally, from the figure, one can see that when the vector coupling is increased the CEP position is pushed to the left, while the coexistence phase becomes wider.

To conclude this section we present our numerical results for the speed of sound, whose square ($c_s^2$) is defined as the slope of $P$ vs. $\mathrm{\varepsilon}$.
This quantity is relevant in the astrophysical applications as it is related with the stiffness of the EOS.
Weak coupling (perturbative) QCD suggests that the conformal limit ($c_s^2  = 1/3$) is approached from below \cite{Cherman:2009tw,Bedaque:2014sqa} (see also Eq. (5) of Ref.~\cite{Kapusta:2021ney}).

\begin{figure}[!ht]
    \centering
    \includegraphics[width=0.48\textwidth]{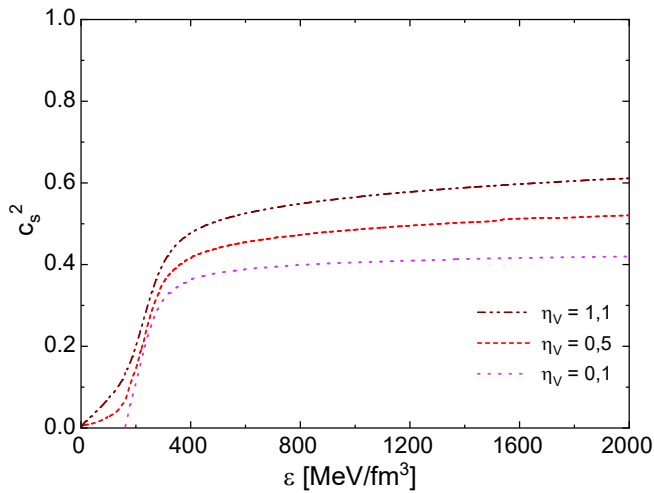}
    \caption{Squared speed of sound for the nonlocal chiral QM model with $\mathrm{\eta_D}=1.1$ and $\mathrm{\eta_V} = 0.1, 0.5$, and $1.1$, respectively.  }
    \label{fig:cs2}
\end{figure}

In Fig.~\ref{fig:cs2} we show the squared speed of sound for fixed diquark coupling ratio $\mathrm{\eta_D} = 1.1$ and for different values of $\mathrm{\eta_V}$, namely, $0.1$, $0.5$, and $1.1$, as function of the energy density $\varepsilon$.
It is clear that, within the present QM model that includes 2SC, $c_s^2$ is always larger than the conjectured limit $1/3$ on QCD, and smaller than $1$, preserving causality. 
In the range of energy densities which is relevant for the cores of neutron stars, $400 < \varepsilon <2000$~${\rm MeV/fm}^3$, and which is displayed in Fig.~\ref{fig:cs2}, the  values for $c_s^2$ are approximately constant and lie in the range $\approx 0.4 - 0.6$, depending on the parameter set.

In order to understand the role of color superconductivity for $c_s^2$, it is instructive to apply formula (6) from Ref.~\cite{Blaschke:2021poc}, which quantifies the deviation from the conformal limit by 
\begin{equation}
\label{eq:cs2}
    c_s^2=\frac{1+\zeta}{3+\zeta}~,~~
    \zeta=\frac{18 \Delta^2}{\xi_4\, a_4\,\mu_B^2}, 
\end{equation}
where $\Delta$ is the pairing gap and $a_4=1-2\alpha_s/\pi$ is the $\mathcal{O}(\alpha_s)$ perturbative correction factor to the ideal massless quark pressure; $\xi_4=1.857$ \cite{Zhang:2020jmb} specifies the application of this formula to the 2SC phase considered here. 
With $\Delta = 150$\,MeV at $\mu_B=1$\,GeV (see Fig.~\ref{fig:VEVs}) and $a_4=0.3$ \cite{Blaschke:2021poc}, we get from Eq.~(\ref{eq:cs2}) the value $c_s^2=0.46$, in rough agreement with the red line in Fig.~\ref{fig:cs2}.
In addition, we have checked that, within our model without interactions ($\mathrm{\eta_D} = \mathrm{\eta_V} = 0$), the behavior of $c_s^2$ corresponds to the conformal limit value of 1/3.

\section{Hybrid EOS}
The aim of the present work is to test the presented QM model in the astrophysical arena.
For that purpose we use a two-phase description to account for the transition from nuclear matter to QM in the interiors of compact stars.

The nonlocal NJL model is found to provide a basic understanding for the mechanisms governing both the spontaneous breakdown of chiral symmetry and the dynamical generation of massive quasiparticles from almost massless current quarks, in close contact with QCD~\cite{Dumm:2021vop}. 

However, it does not account for some important features expected from the underlying QCD interactions. 
In particular, the model predicts the existence of colored quasiparticles in regions of $T$ and $\mu$ where they should be suppressed by confinement.
Therefore, to successfully describe the dynamics of QCD, it is necessary to include the effects of color confinement, that could be mimicked through a bag pressure term~\cite{Blaschke:2010vj}.
In the hybrid description we observed that both, hadronic and QM, EOS have similar behavior in the neighborhood of the phase transition which has been named as the masquerade effect \cite{Alford:2004pf} )see also \cite{Blaschke:2007ri,Klahn:2006iw,Blaschke:2006gd}).
Then, below, we will show that to satisfactory fulfill modern astrophysical constraints for the neutron star masses and tidal deformabilities within the present hybrid description, the EOS (total pressure) of the nonlocal NJL model should be softened (modified) near the phase transition (from hadronic to QM phase) in order to avoid the masquerade effect, and stiffened in the high-density range to reach the maximum mass prediction for compact stars. 
A fruitful way to solve both drawbacks at zero temperature is to include a bag pressure in the total pressure of the QM description.
Therefore, we will start by considering a constant bag scenario in Sec.~\ref{sectionBconst} and then, in Sec.~\ref{sectionBmu}, we will generalize this approach by a $\mu_B$-dependent bag pressure.

\subsection{Hadronic model}
To describe nuclear matter the relativistic density-functional approach by Typel \cite{Typel:2005ba} has been used, which includes meson-exchange interactions within the  “DD2” parametrization \cite{Typel:2009sy}.
Even though this model satisfactorily describes nuclear matter up to the saturation density, at higher densities the above-mentioned masquerade problem occurs for the hybrid construction.
Therefore, we explore the hadronic EOS including a recent reformulation of the excluded volume effect~\cite{Typel:2016srf} to mimic a repulsive interaction resulting in a proper phase transition.
This nucleonic excluded volume has a microscopic background. It emulates the quark Pauli blocking effect among nucleons as a necessary consequence of their quark substructure and the overlap of the nucleon wave functions.
In the present work, different values of the excluded volume were considered, for instance, DD2-p40.
Here, the label “p40” stands for a positive excluded volume parameter of $v = 4\,\text{fm}^3$.
This type of nuclear EOS has been extensively used in systematic studies of hybrid star models (see Refs.~\cite{Benic:2014jia, Kaltenborn:2017hus, Alvarez-Castillo:2016oln, Alvarez-Castillo:2018pve}).
In addition, it has also been considered in a neutron star crust, including the Baym-Pethick-Sutherland (BPS) model \cite{Baym:1971pw}, to fully describe the hadronic EOS at low baryon densities.

\subsection{Maxwell construction}
The present work only considers a sharp interphase between QM and hadronic phase.
That means that no mixed phase is expected (see Ref.~\cite{Lugones:2021tee} and references therein).
Then, the phase transition between the EOS for nuclear matter and QM will be described by the Maxwell construction.
Both phases satisfy tha charge-neutrality condition and $\beta$ equilibrium with electrons and muons. 
Then, both phases are connected  by requiring that chemical potentials and pressures of the two phases coincide at the phase transition,
\begin{equation}
\mu^H = \mu^{QM} = \mu_c   
\end{equation}
and 
\begin{equation}
P^H = P^{QM} = P_c.   
\end{equation}
Outside the phase transition, the phase with higher pressure (lower grand canonical potential) is to be chosen as the physical one.

It is worth mentioning that the first result we obtained (not shown here) is that unless additional (de)confining effects are introduced with a bag pressure, the only physical crossings between hadronic and QM EOS occur at very high $P$ and $\mu_B$ values, producing compact star masses, radii, and tidal deformabilities that do not accomplish several astrophysical constraints mentioned in the Introduction (see Sec.~\ref{sect:intro}).

\subsection{Calculation of astrophysical observables}
In order to rate and compare the obtained EOS  we have to calculate possible neutron star properties. 
Therefore, to generate plausible solutions for neutron star properties, hybrid neutron star EOoS were augmented with the crust EOS by Baym, Pethick, and Sutherland \cite{Baym:1971pw}.
In the first step one can calculate the range of possible neutron star radii and masses. 
These can directly be compared to the combined observations from NICER and XMM Newton of the millisecond pulsar J0740+6620 according to the analysis of Miller {\it{et al.}}\,\cite{Miller:2021qha}.
To evaluate them one has to solve the TOV equations for a static nonrotating, spherical-symmetric star \cite{PhysRev.55.364,PhysRev.55.374} (see Appendix \ref{sect:TOV_tidals} for details).
The tidal deformability $\Lambda$ can be calculated for the considered sequence of neutron star masses\,\cite{Hinderer:2009ca} and compared to the constraint obtained from the gravitational wave signal that was observed for the binary neutron star merger GW170818\,\cite{LIGOScientific:2018cki} in the mass range $M\approx 1.4M_\odot$. 

The astrophysical observables were calculated on the basis of a code by Maselli\,\cite{maselli2021}.
\\

\subsection{Constant bag pressure}
\label{sectionBconst}

As a first step towards astrophysical applications, we consider a shift of the QM EOS by a constant bag pressure 
$B(\mu_B)={\rm const}=B$ that allows to lower the crossings between the hadronic and QM curves in the $P$ vs $\mu_B$ plane (Maxwell construction). 

\begin{widetext}

\begin{figure}[H]
    \centering
    \includegraphics[width=0.45\textwidth]{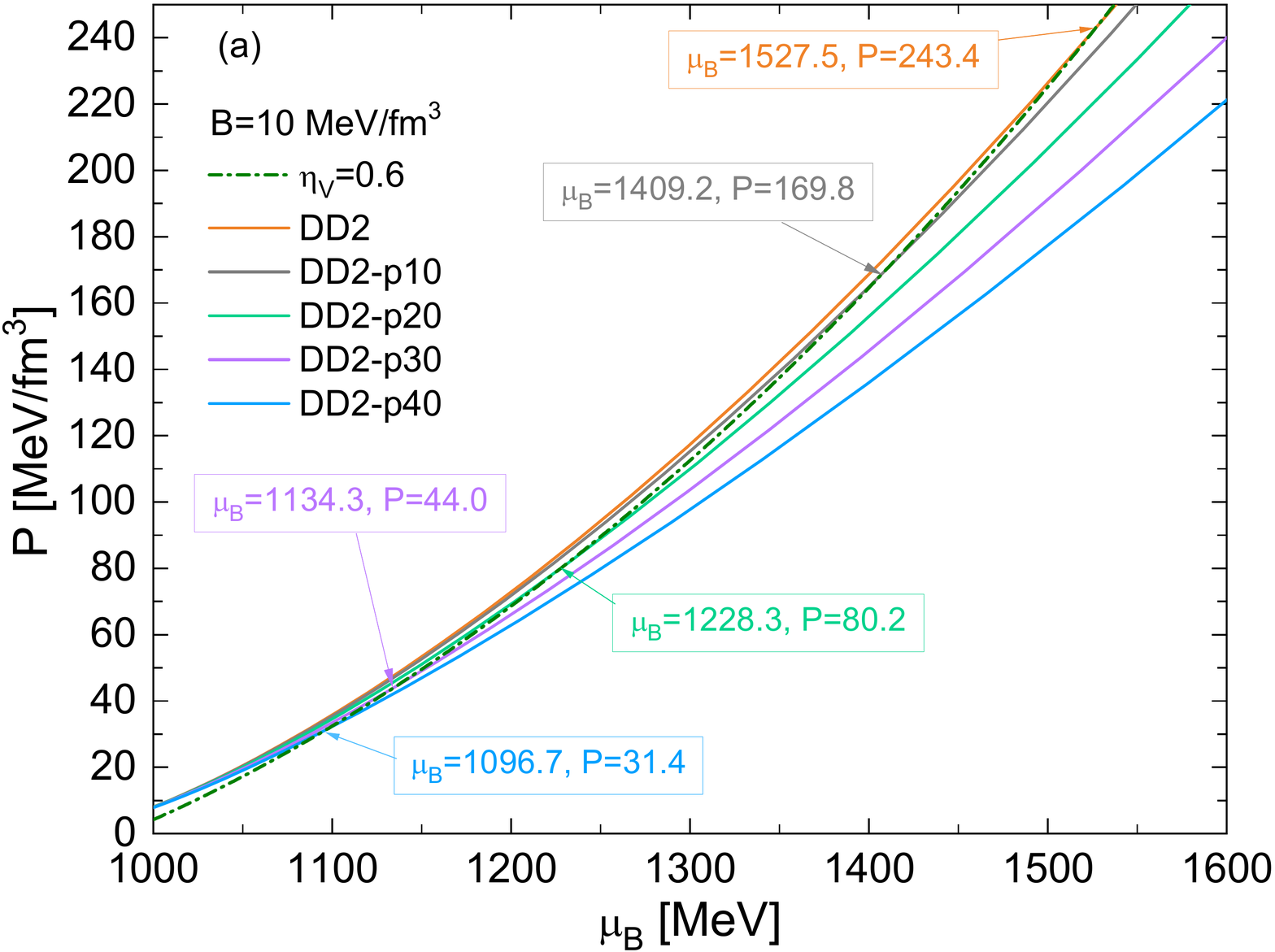}
    \includegraphics[width=0.45\textwidth]{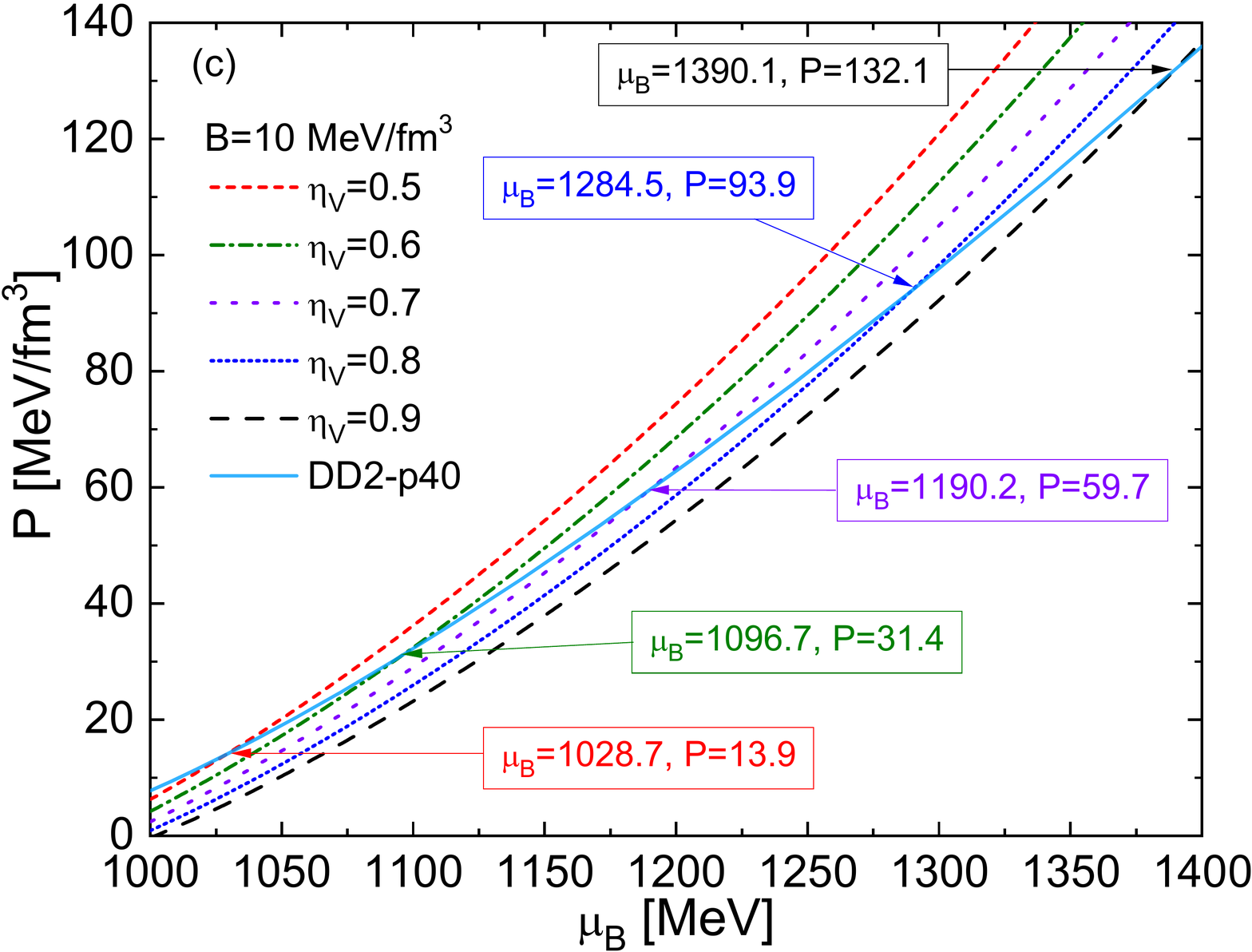} \\    
    \includegraphics[width=0.45\textwidth]{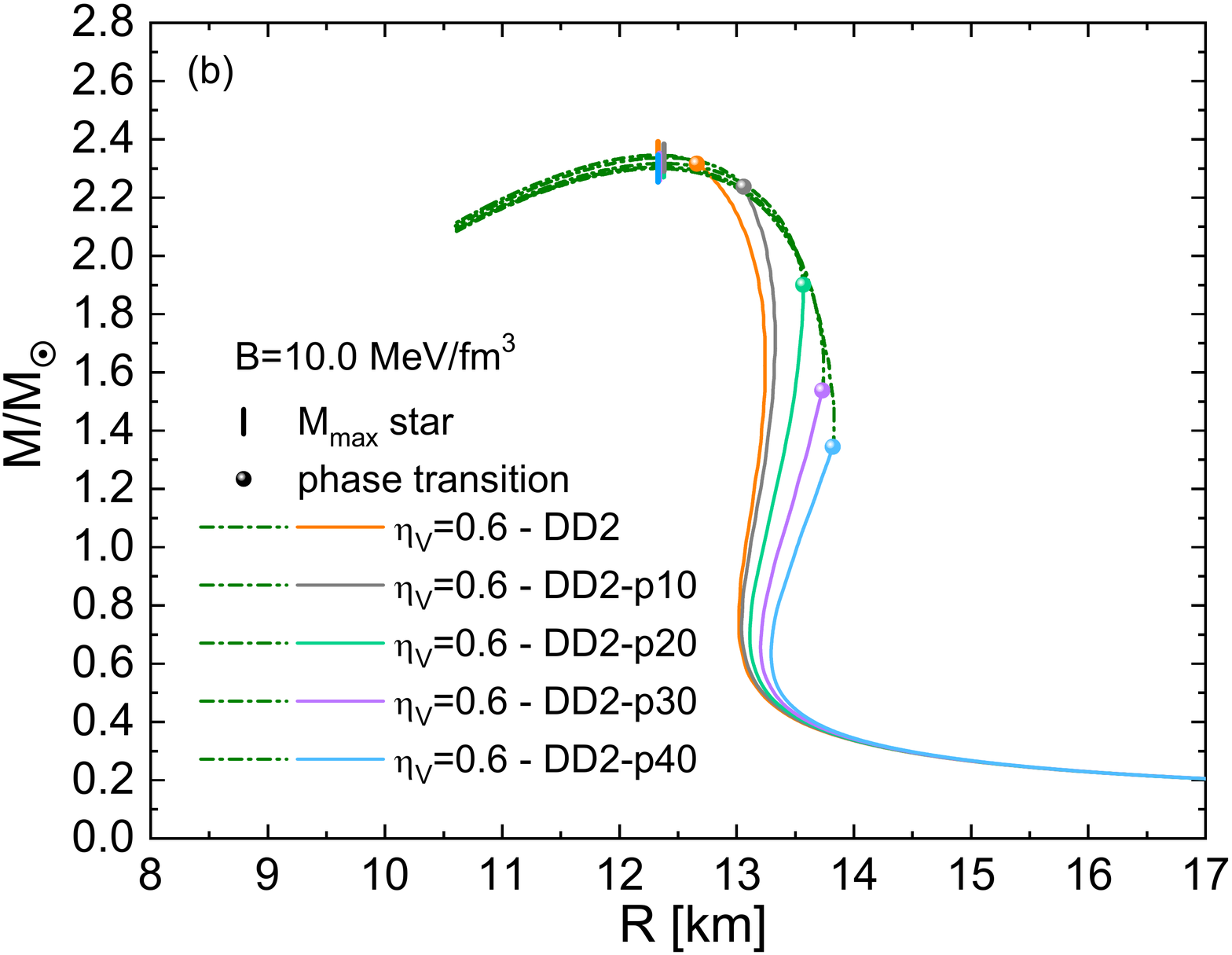}
    \includegraphics[width=0.45\textwidth]{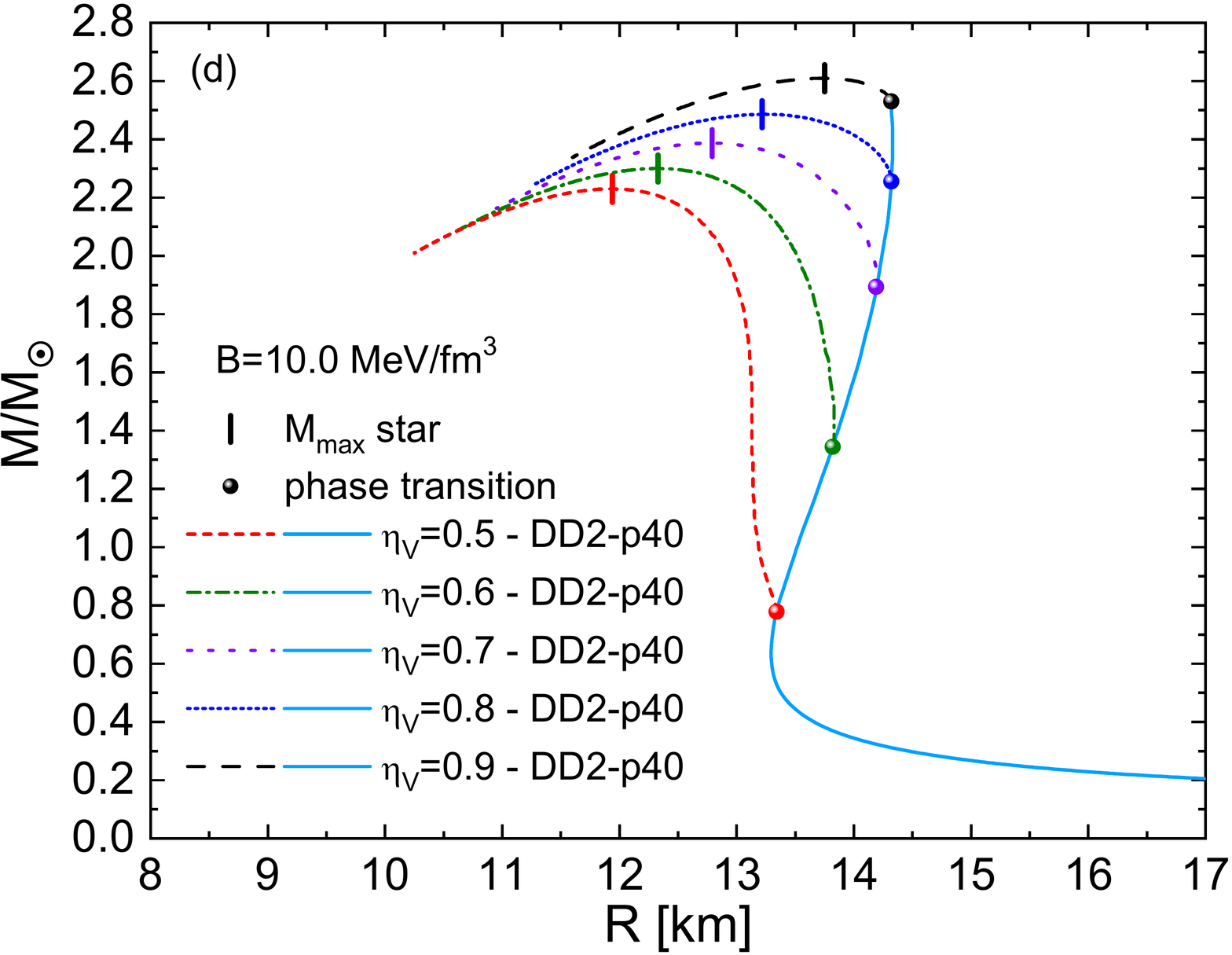}
    \caption{Left: (a) QM EOS for $\eta_D=1.1$ and Bag = 10.0 MeV/fm$^3$ for $\eta_V=0.6$ (green dash-dotted line), and different hadronic EOS with solid lines. (b) The corresponding $M$-$R$ relations for the hybrid EOS in (a). The solid dots indicate the respective phase transition points and vertical bars the location of the maximum mass (down). Right: (c) QM EOS with $\eta_D=1.1$, Bag = 10.0 MeV/fm$^3$, and $\eta_V$ from 0.5 to 0.9 (different trace lines), and hadronic EOS DD2-40 in light blue solid line. (d) The corresponding $M$-$R$ relation.}
    \label{fig:etaD_1.1_B10_0.6-all_all-DD2p40}
\end{figure}

\end{widetext}

We renormalize the QM EOS considering that
\begin{equation}
P(\mu_B) \rightarrow  P(\mu_B) - B,    
\end{equation}
where a bag pressure shift is included in the QM EOS. Throughout this section, we will consider a fixed $B = 10$ MeV/fm$^3$ that lies in the range 10-50  MeV/fm$^3$ used in Ref. \cite{Blaschke:2010vj}.
With this particular choice we are able to obtained an early onset from hadronic to QM phase as can be see from Fig.~\ref{fig:etaD_1.1_B10_0.6-all_all-DD2p40}.

First, in order to explore the effects of setting different parameter values, for example, the $\eta_V$ values for QM and the excluded volume parameter for the hadronic phase, we proceed as follows: First we consider a hybrid description for a fixed QM EOS and different hadronic ones.
Later we choose one hadronic EOS and different QM EOS.
After that, we will only present some representative hybrid EOS. 

In Fig.~\ref{fig:etaD_1.1_B10_0.6-all_all-DD2p40}(a) we show a particular QM EOS for fixed $\eta_D=1.1$, $\eta_V=0.6$ (green dashed line), and different hadronic EOS: DD2 (orange) or DD2-p10 (black), p40 (light blue) and some values in between, p20 (green), and p30 (purple).
As expected, the stiffer the hadronic EOS, the earlier the crossing with the QM EOS.
These EOS were considered as input for solving the TOV equations, giving as output the $M$ vs $R$ shown in Fig.~\ref{fig:etaD_1.1_B10_0.6-all_all-DD2p40}(b).
Earlier onsets can be observed for larger excluded volume parameters in the hadronic EOS, corresponding to stiffer hadronic EOS.
In all cases, the maximum mass attained is about 2.3$M_{\odot}$ with  $R \approx$ 12.35 km.

However, in Fig.~\ref{fig:etaD_1.1_B10_0.6-all_all-DD2p40}(c) we show one hadronic EOS, DD2-p40 (solid line), and different QM EOS with $\eta_D=1.1$ and $\eta_V$ from 0.5 to 0.9 (different line styles).
In Fig.~\ref{fig:etaD_1.1_B10_0.6-all_all-DD2p40}(d), we show the $M$-$R$ relations that correspond to the parameter selections shown in Fig.~\ref{fig:etaD_1.1_B10_0.6-all_all-DD2p40}(c).
One observes that the increase of $\eta_V$ shifts the onset of deconfinement to higher masses.
At the same time, the maximum mass increases with $\eta_V$, exceeding even the value of about 2.6$M_{\odot}$ attained at $R \approx 13.75$~km for $\eta_V$ = 0.9.
Since the vector interactions produce a stiffer QM EOS, the larger the vector coupling, the larger both maximum mass and radius, but at the price of a later onset of deconfinement.

\begin{widetext}

\begin{figure}[H]
    \centering
    \includegraphics[width=0.47\textwidth]{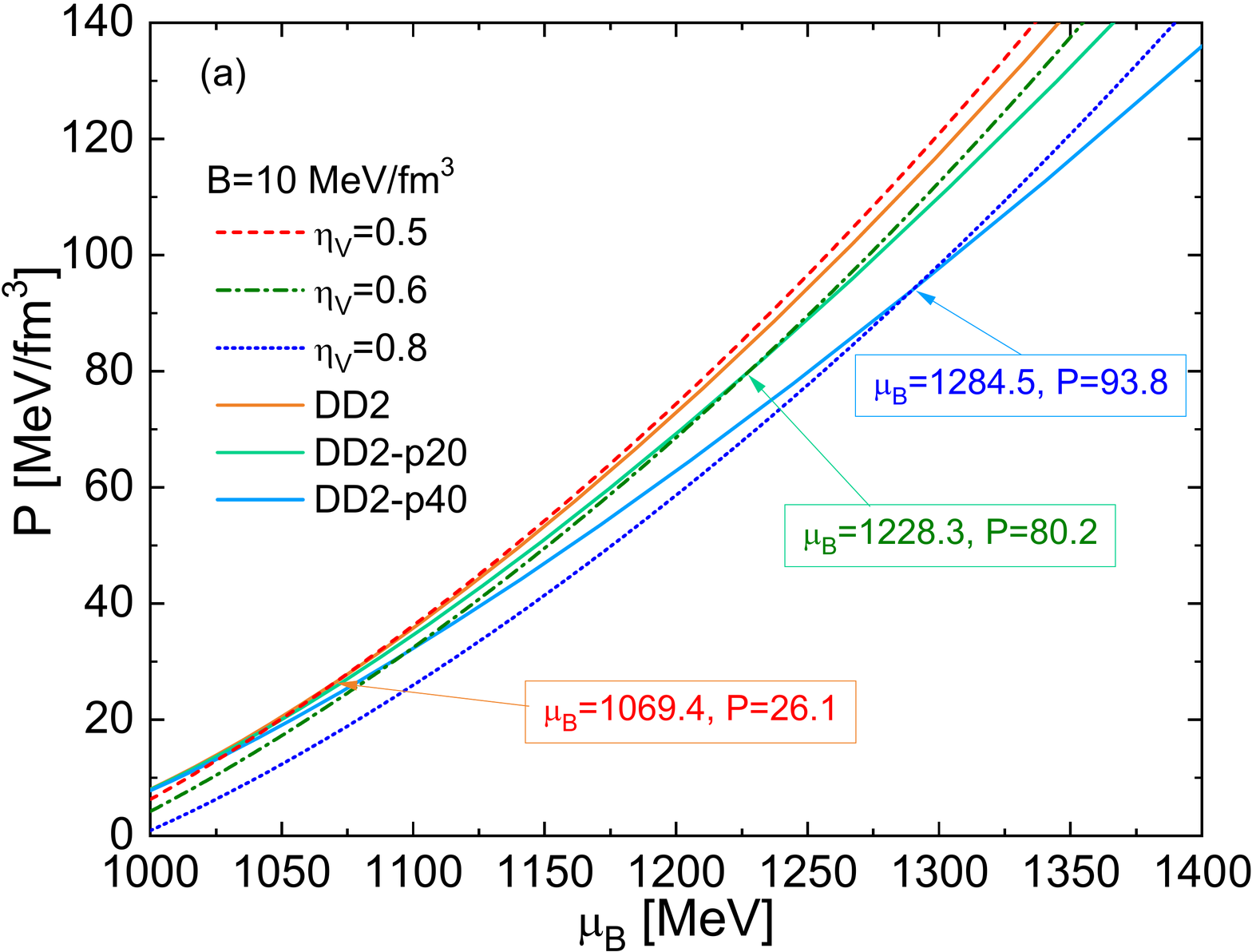}
    \includegraphics[width=0.47\textwidth]{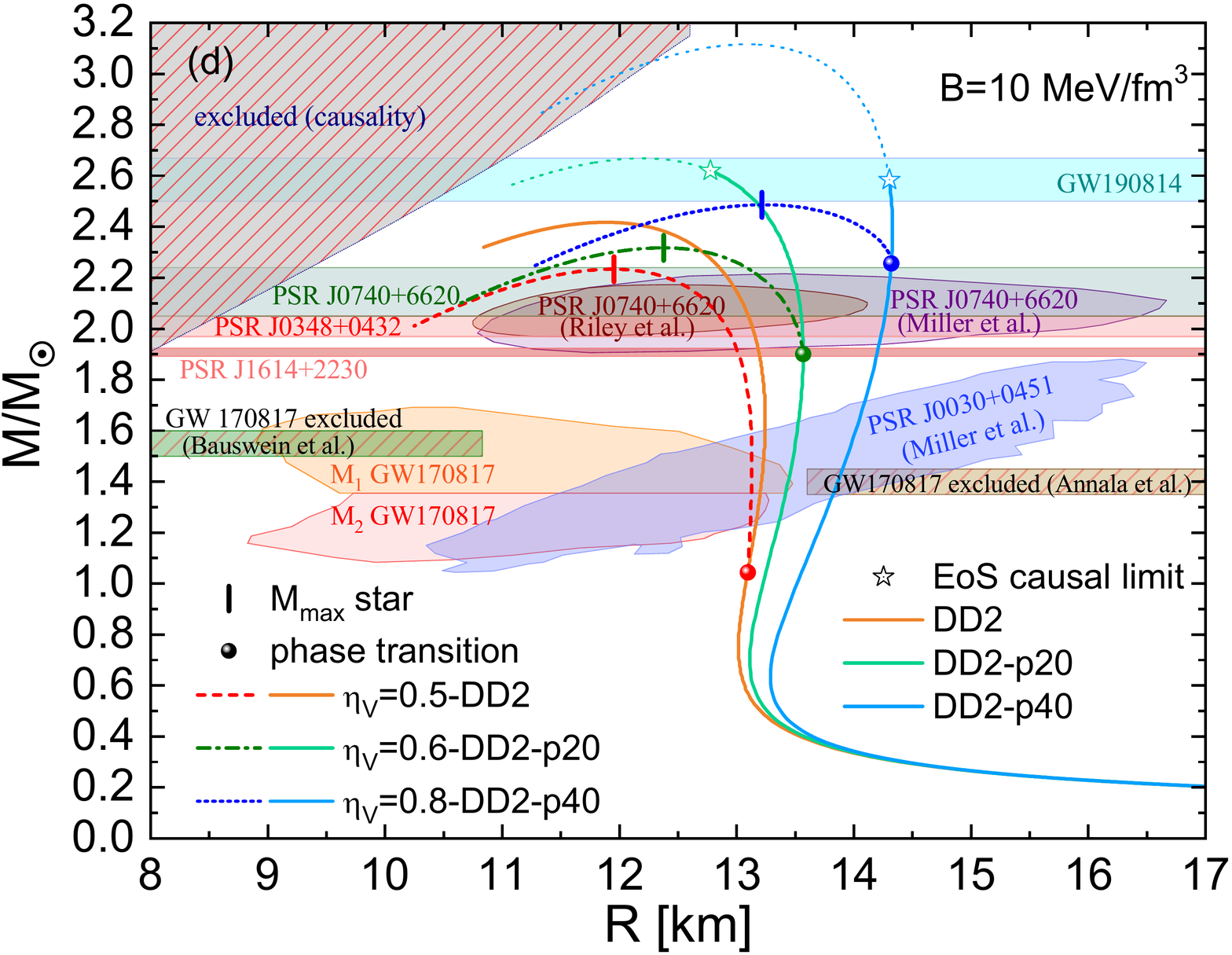} \\
    \includegraphics[width=0.47\textwidth]{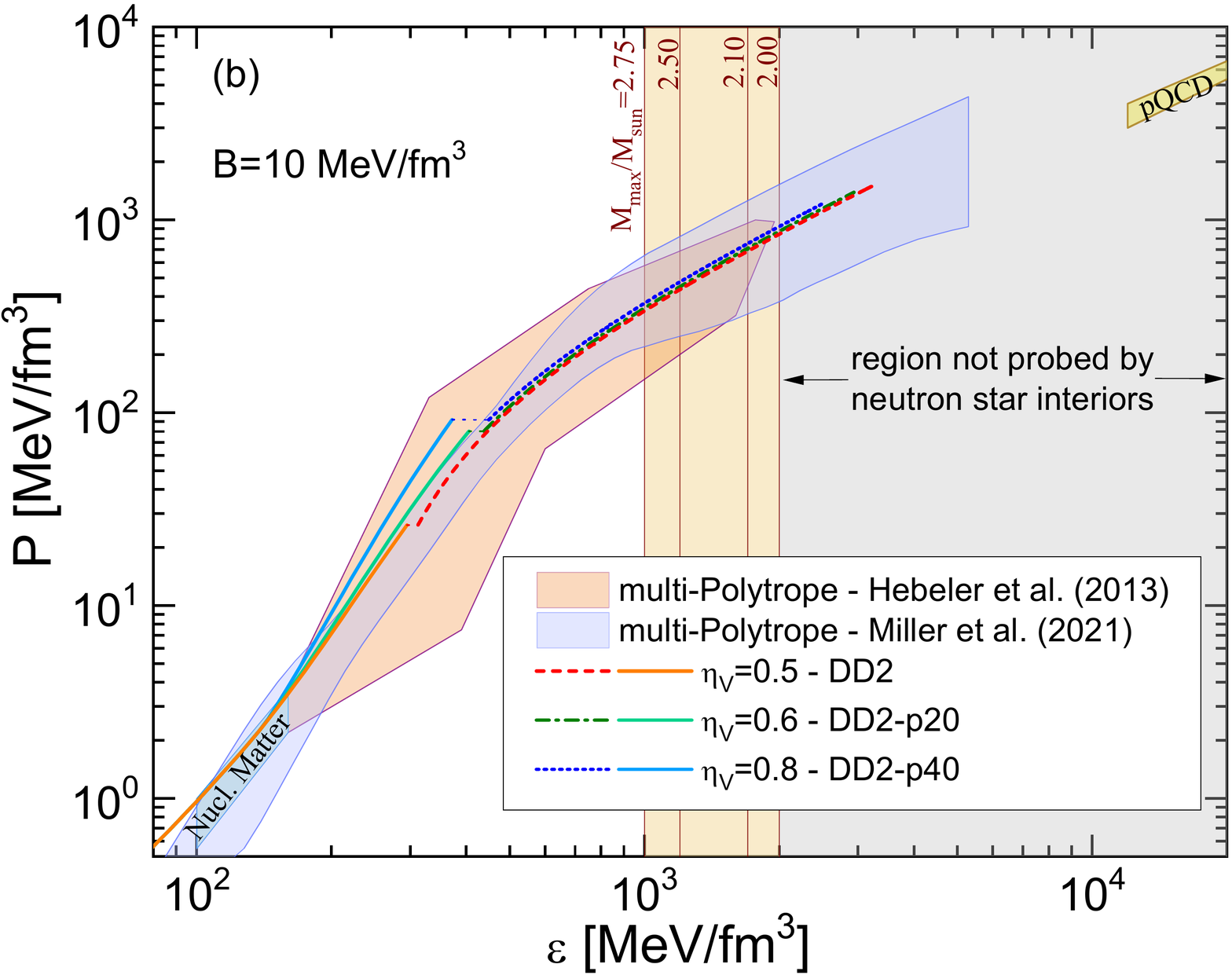}
    \includegraphics[width=0.47\textwidth]{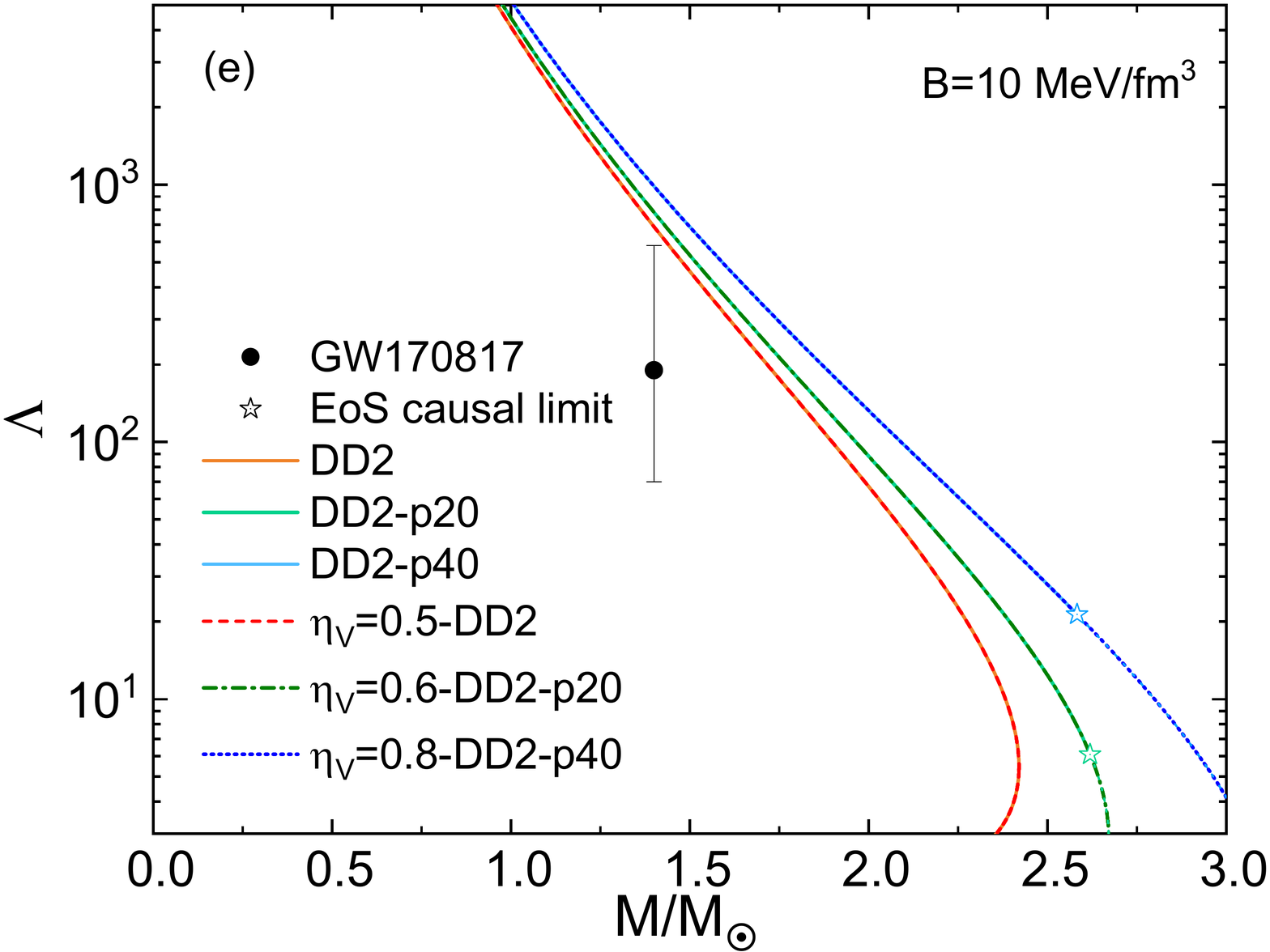} \\
    \includegraphics[width=0.47\textwidth]{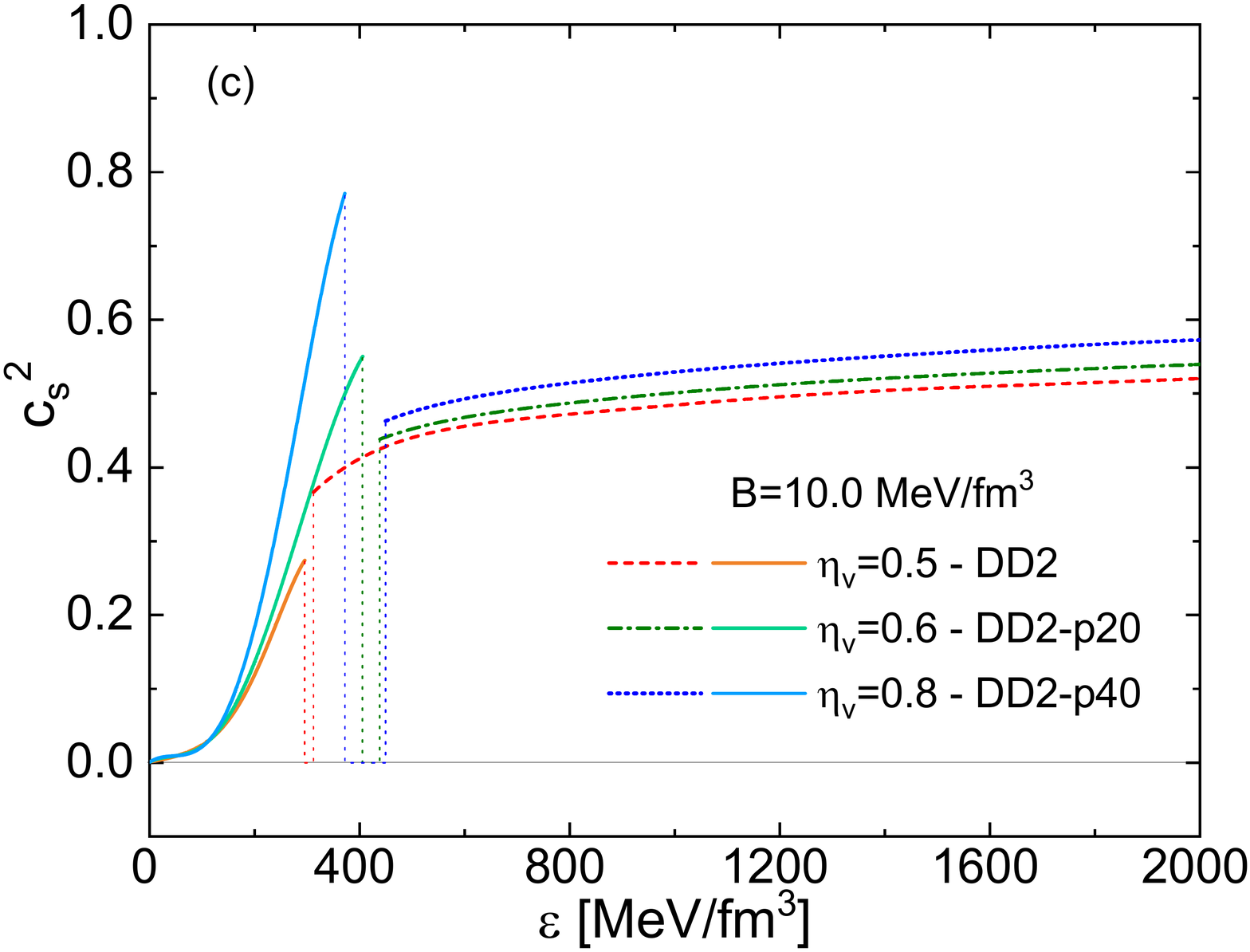}
    \includegraphics[width=0.47\textwidth]{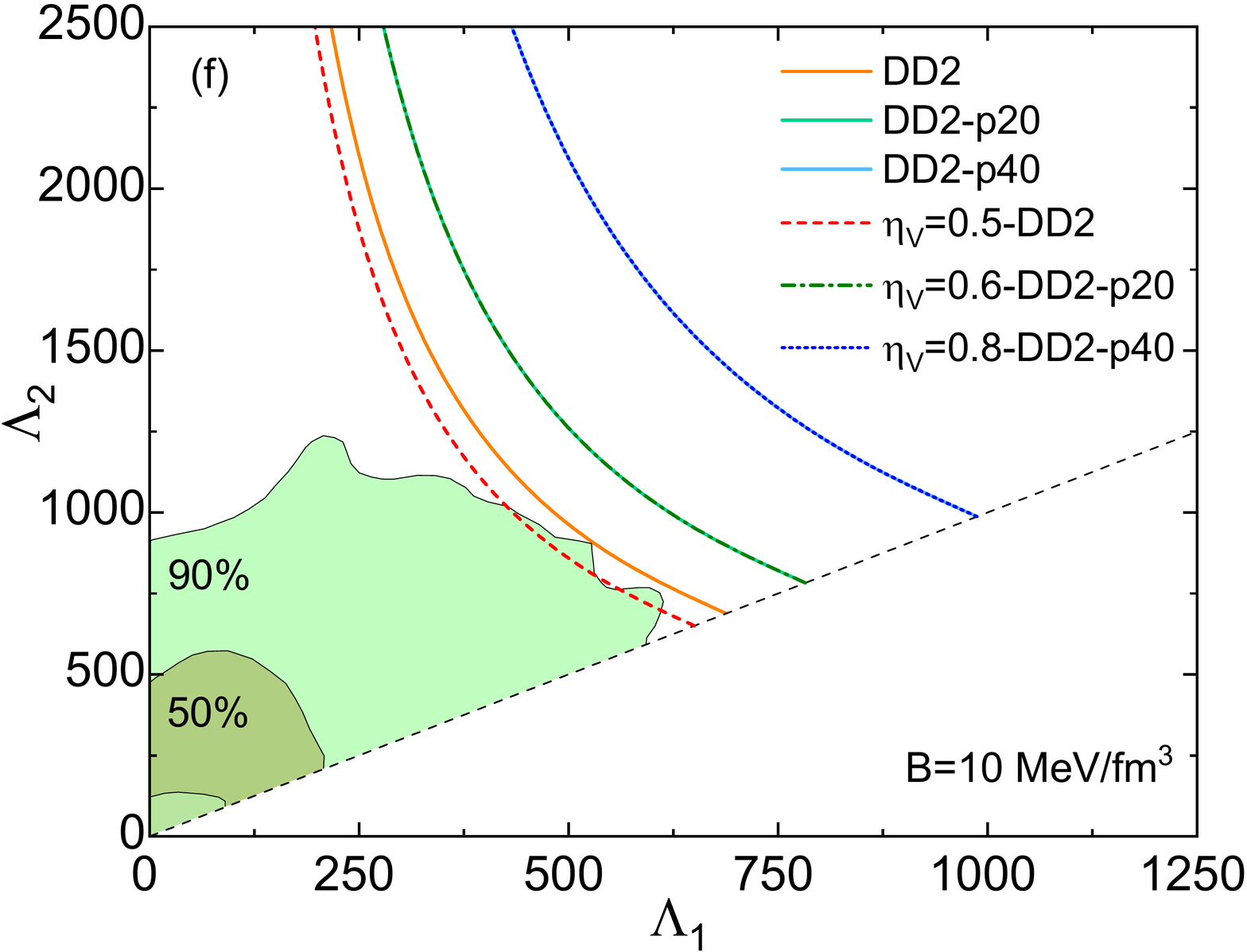}
    \caption{\small{(a) Pressure vs chemical potential for the QM (dashed lines) and the hadronic (solid lines) EOS. (b) Pressure vs energy density, and we highlight the fact that NS interiors do not probe the energy densities where pQCD is applicable. (c) The squared sound speed $c_s^2$ is shown for the corresponding hybrid EOS. (d) The $M$-$R$ relations are given. The solid dots indicate the respective phase transition points and vertical bars show the location of the maximum mass stars. The results for the pure hadronic EOS are also shown (lighter line colors), with the star symbols indicating the points where the causal limit of the EOS is reached and the dotted lines stand for results from its acausal part. For comparison with the NS phenomenology, the mass-radius constraints from NICER  observations (see Refs. \cite{Miller:2021qha,Miller:2019cac} for details) and from the merger event GW170817 are shown. (e) The tidal deformability $\Lambda$ vs $M/M_{\odot}$ for the set of hybrid EOS (traced lines), hadronic EOS (solid lines, overlapped with QM ones), and comparison to the results of GW170817. (f) The comparison of $\Lambda_2-\Lambda_1$ calculated for our EOS in comparison to the analysis of the gravitational wave signal from the merger GW170817 \cite{LIGOScientific:2017vwq, De:2018uhw}. For the remaining mass and radius constraints in (d) see Ref.~\cite{Ayriyan:2021prr}.}}
    \label{fig:all_B10}
\end{figure}

\end{widetext}

In Fig.~\ref{fig:all_B10} we summarize the results for the hybrid EOS and their relation to the NS observational constraints.
Figure~\ref{fig:all_B10}(a) shows the Maxwell constructions for three hybrid EOS cases: an early phase transition onset ($\eta_V=0.5$, DD2, red lines), a high mass onset ($\eta_V=0.8$, DD2-p40, blue lines), and an intermediate onset value ($\eta_V=0.6$, DD2-p20, green lines). 

In Fig.~\ref{fig:all_B10}(b) we compare these hybrid EOS with a well-known constrained region from Hebeler {\it{et al.}} \cite{Hebeler:2013nza} that accommodates all EOS discussed before.
The comparison with the confidence region of the Bayesian analysis for a multiparameter EOS model (piecewise polytrope in this case) with the NICER and X-ray Multi-Mirror (XMM-Newton) observations of PSR J0740+6620 (see Ref.~\cite{Miller:2021qha} for details) shows a very good agreement for all the presented EOS with the only exception of DD2-p40 which has a too late onset of deconfinement for the very stiff hadronic phase. 
A similar discussion of the admissible EOS region can be found  in Ref.  \cite{Annala:2019puf} based on a large number of individual EOS generated by a speed-of-sound interpolation method.

In Fig.~\ref{fig:all_B10}(c) the squared speed of sound is shown for the hybrid EOS from Fig.~\ref{fig:all_B10}(a). The behavior of $c_s^2$ in the quark matter phase is rather constant in a narrow range around $\approx 0.5$, for $\eta_V$ = 0.5-0.8.
We note that in Ref.~\cite{Antic:2021zbn} the reverse observation has been made. Namely, a constant speed of sound model for quark matter provides a perfect fit to a nonlocal NJL model with covariant form factors.

The $M$-$R$ plot is presented in Fig.~\ref{fig:all_B10}(d), where we show the hybrid solutions together with the purely hadronic ones. As in previous very recent works \cite{Ayriyan:2021prr, Cierniak:2021vlf, Ivanytskyi:2021dgq}, we show 
for comparison different colored regions corresponding to either pulsar measurements or forbidden (striped) regions that serve as constraints for the compact star EOS. 
The green band region above 2$M_{\odot}$ corresponds to the updated mass measurement of PSR J0740+6620~\cite{Fonseca:2021wxt}, which was recently upgraded to a mass-radius measurement by NICER, shown by the purple ellipsoidal type region for the result of the Maryland-Illinois team~\cite{Miller:2021qha} and by the wine colored region for XMM-Newton Spectroscopy~\cite{Riley:2021pdl}. 
The orange and red regions around $M =1.4M_{\odot}$ correspond to the compactness estimates for the components labeled as $M_1$ and $M_2$ of the binary NS merger GW170817 obtained from the gravitational wave signal of its inspiral phase~\cite{LIGOScientific:2018cki}.
The central big blue region corresponds to the $95\%$ contour of the joint probability density distribution of mass and radius from NICER observation of PSR J0030+0451 \cite{Miller:2019cac}.
The green and brown striped bands correspond to excluded regions derived from GW170817 observations by  Bauswein {\it{et al.}}~\cite{Bauswein:2017vtn} and Annala {\it{et al.}}~\cite{Annala:2017llu}.
The light blue band region corresponds to the mass $2.59^{+0.08}_{-0.09}~M_{\odot}$ of the lighter component in the binary merger event GW190814 \cite{LIGOScientific:2020zkf}.
Older maximum mass constraints from Shapiro Delay measurement of PSR J0348+0432 and PSR J1614+2230 \cite{Demorest:2010bx,Lynch:2012vv,Antoniadis:2013pzd,Fonseca:2016tux,Arzoumanian_2018} are also included as additional references.

In Fig.~\ref{fig:all_B10}(e) we show the tidal deformability as a function of $M/M_{\odot}$ including its value ($\Lambda=190^{+390}_{120}$) from GW170817  \cite{LIGOScientific:2018cki}.
Finally, in Fig.~\ref{fig:all_B10}(f) we present the tidal deformability parameters $\Lambda_1$ and $\Lambda_2$ of the high and low mass components of the binary merger (see Appendix \ref{sect:TOV_tidals} for details) compared to the probability density contours for the analysis of GW170817 signals \cite{LIGOScientific:2017vwq, De:2018uhw}.
The $\Lambda_1$ and $\Lambda_2$ parameters characterize the size of the tidally induced mass deformations of each star.

Note that for tidal deformability results, the hybrid EOS are not separated into their QM and hadronic parts; then in Figs.~\ref{fig:all_B10}(e) and \ref{fig:all_B10}(f) the traced lines represent all the hybrid EOS and the solid lines are for the pure hadronic EOS as usual, but their results are indistinguishable in almost all the cases. 
 
From the results shown in Figs.~\ref{fig:all_B10}(e) and \ref{fig:all_B10}(f) it is evident that even for the case of early onset of deconfinement the hybrid EOS model with a constant bag pressure is not sufficient to fulfill both constraints, maximum mass and tidal deformabilities, simultaneously.
Therefore a more general approach is necessary.

\subsection{Density-dependent bag pressure}
\label{sectionBmu}

There is a way out that goes beyond the standard formulation of the nonlocal NJL-type models by including density-dependent coefficients.
That approach promotes the nonlocal NJL relativistic mean field setting to a model with medium-dependent coupling ($\mu_B$ dependence).
We were inspired by Ref.~\cite{Alvarez-Castillo:2018pve}, where a $\mu_B$ dependence of the bag pressure was included in a covariant nonlocal NJL-type model of quark matter so that the EOS of the density functional approach developed by Kaltenborn {\it{et al.}}~\cite{Kaltenborn:2017hus} could be approximately reproduced.
In our present work, we use the same functional form of the $\mu_B$-dependent bag pressure from Ref.~\cite{Alvarez-Castillo:2018pve}. 
In this way it is possible to describe both an early onset of quark matter and an increase of the maximum mass. This method has been used before, e.g., in Ref. \cite{Maieron:2004af} and in Ref. \cite{Shahrbaf:2019vtf} for solving the hyperon puzzle by early quark deconfinement.
Here we have a similar situation. We need an early onset of the strong phase transition in order to simultaneously have a sufficient compactification and to keep the high value for the maximum mass.
The slope of $B(\mu_B)$ defines a bigger and earlier jump in energy density at the first-order phase transition, which is the necessary change that leads to the compactification of the hybrid stars that allows to fulfill the tidal deformability constraint.

We want to point out that one way to justify a medium dependence of the coupling constants and background pressure in chiral quark models of the NJL type with current-current interactions is to postulate a relativistic density functional
\cite{Kaltenborn:2017hus,Ivanytskyi:2021dgq,Ivanytskyi:2022oxv}. 
Upon expanding around the stationary point, the density-functional approach assumes the form of an effective NJL-type model with medium-dependent couplings (see \cite{Ivanytskyi:2021dgq,Ivanytskyi:2022oxv}). 
In this way it was possible to obtain a strong first-order phase transition with a large enough latent heat of the transition to allow for a separate third family of hybrid stars associated with the phenomenon of mass twin neutron stars~\cite{Alvarez-Castillo:2018pve}.

Then, in what follows we will consider a $\mu_B$-dependent bag pressure as in Ref. \cite{Alvarez-Castillo:2018pve}, given by the equation 
\begin{equation}
B(\mu_B) = B_0 \, f_< (\mu_B)    
\end{equation}
with
\begin{equation}
f_< (\mu_B) = \frac{1}{2}\left[1 - \mathrm{tanh} \left( \frac{\mu_B - \mu_<}{\Gamma_<} \right)\right],  \end{equation}
where we have set $\mu_< = 895$ MeV, $\Gamma_< = 180$ MeV and $B_0 = 35$ MeV/fm$^3$ as the optimal values to reproduce the astrophysical observables. Note that this bag pressure, in the vicinity of the transition from hadronic to QM phase, takes a value around the one considered in Sec.~\ref{sectionBconst}.
Also, it is important to remark that a $\mu_B$-dependent bag pressure will affect the value of $n_B={\partial P}/{\partial \mu_B}$, producing a noticeable effect on the Gibbs free energy via Eq.~(\ref{eq:Gibbs_energy}) and therefore also on the energy density.
This effect is not present in the scenario with $B={\rm const}$.

\begin{widetext}

\begin{figure}[H]
    \centering
    \includegraphics[width=0.49\textwidth]{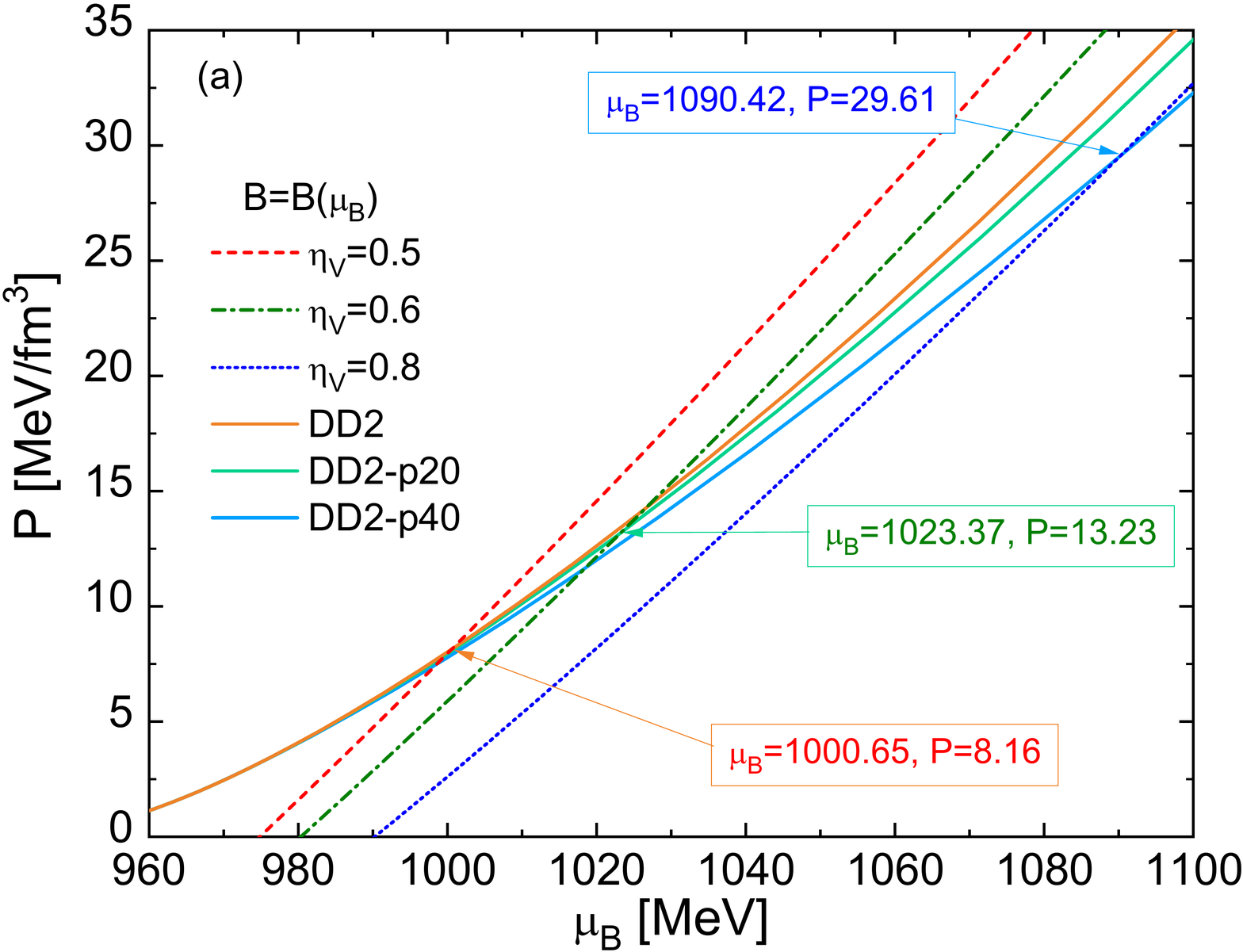}
    \includegraphics[width=0.49\textwidth]{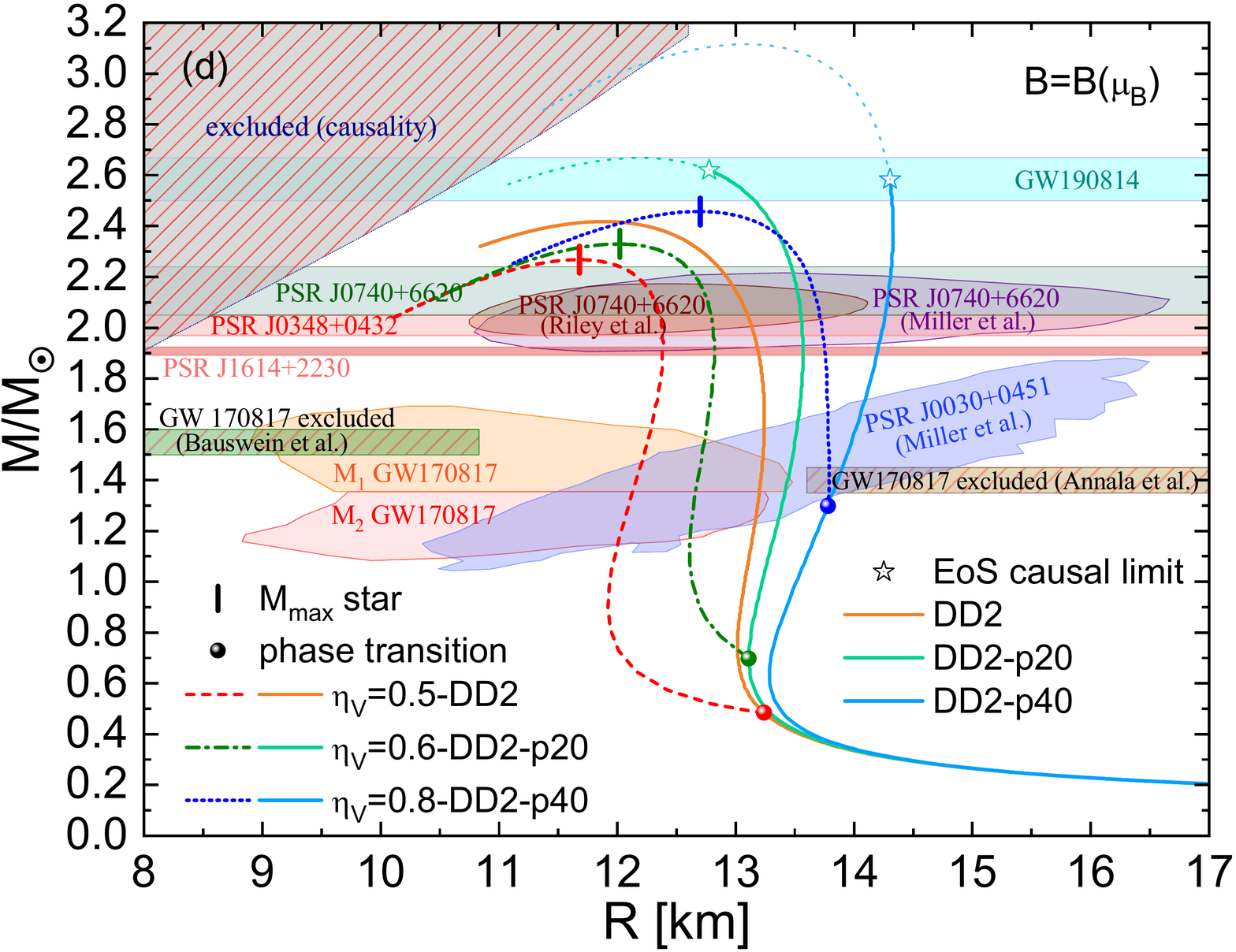} \\
    \includegraphics[width=0.49\textwidth]{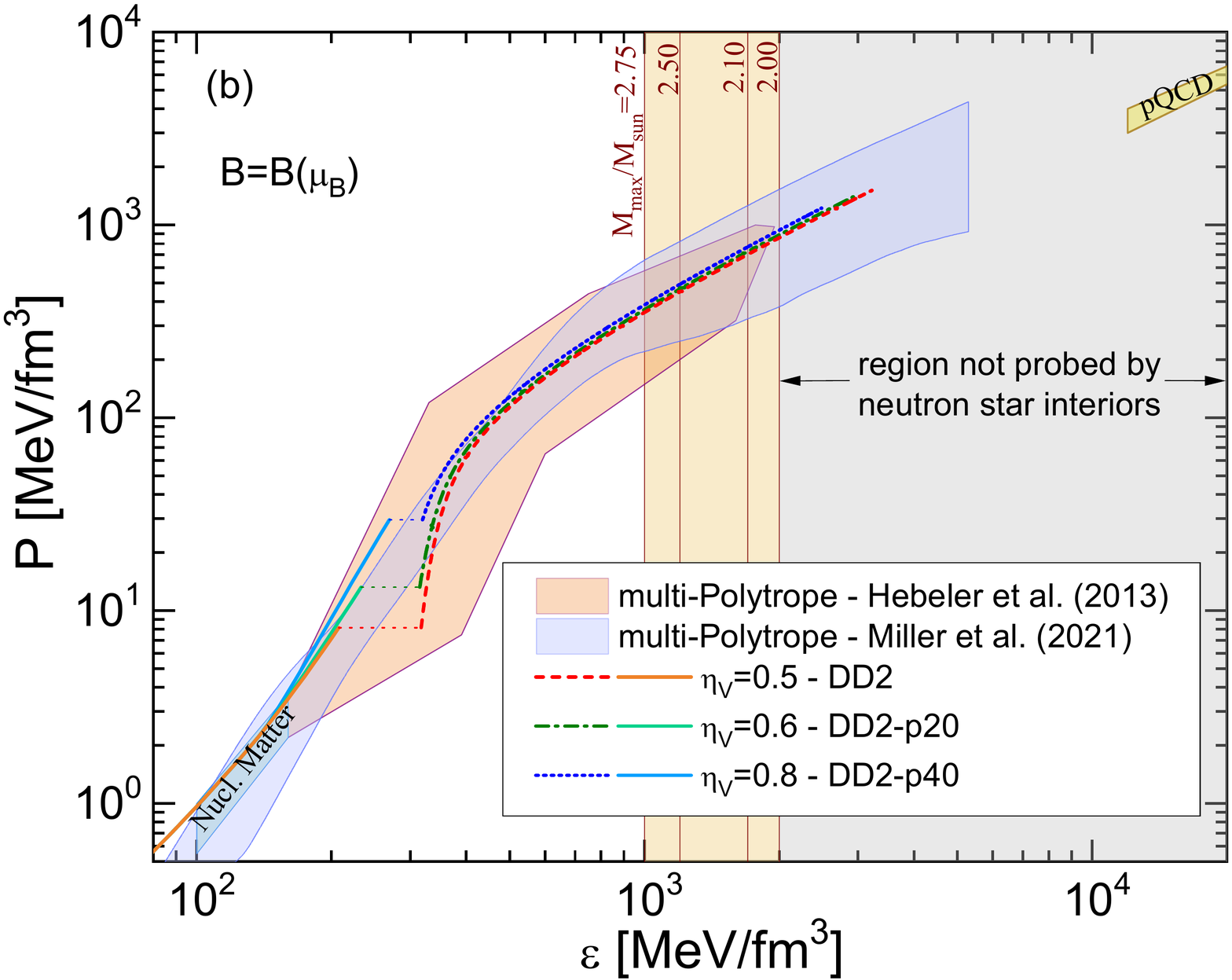}
    \includegraphics[width=0.49\textwidth]{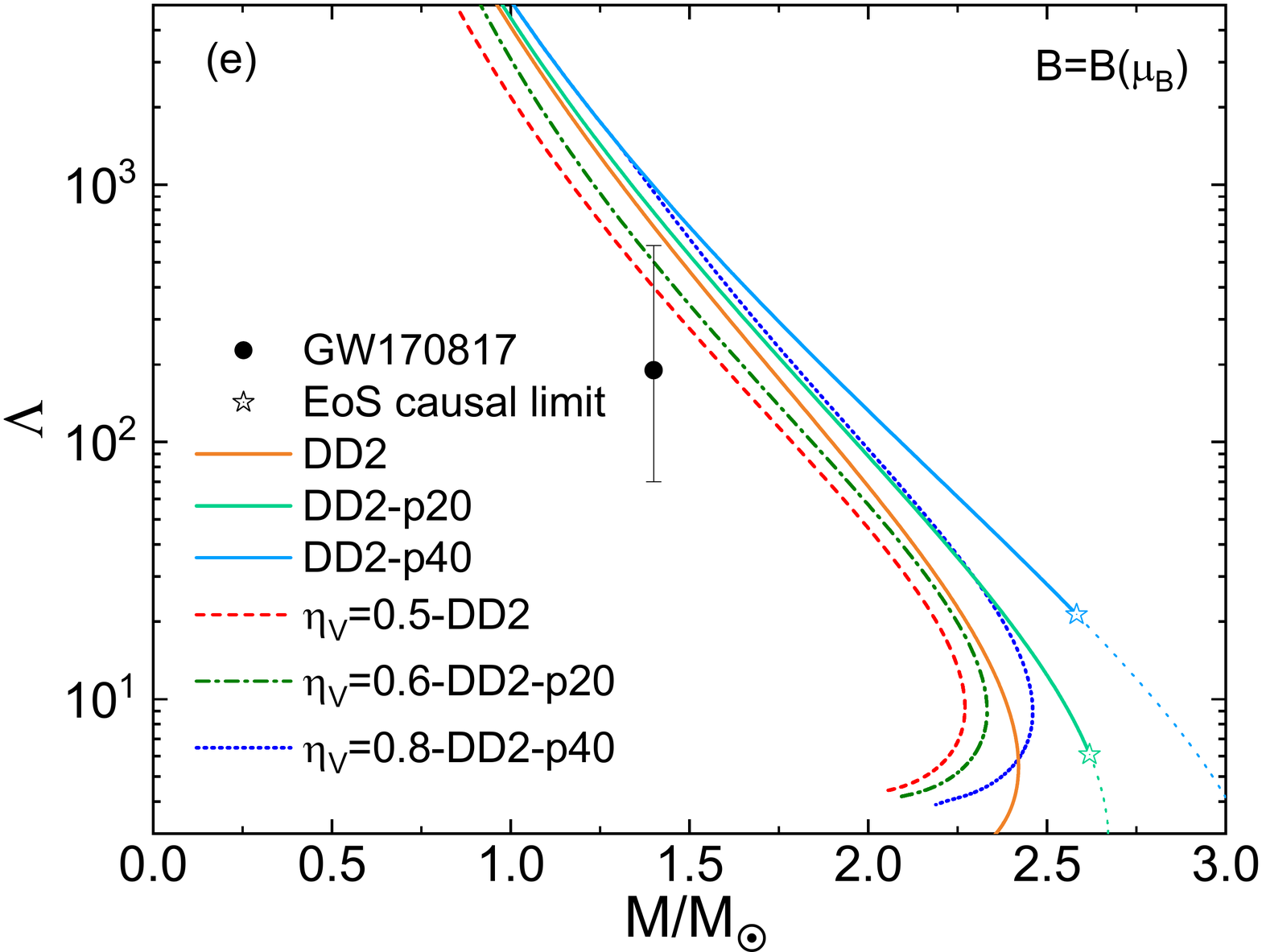} \\
    \includegraphics[width=0.49\textwidth]{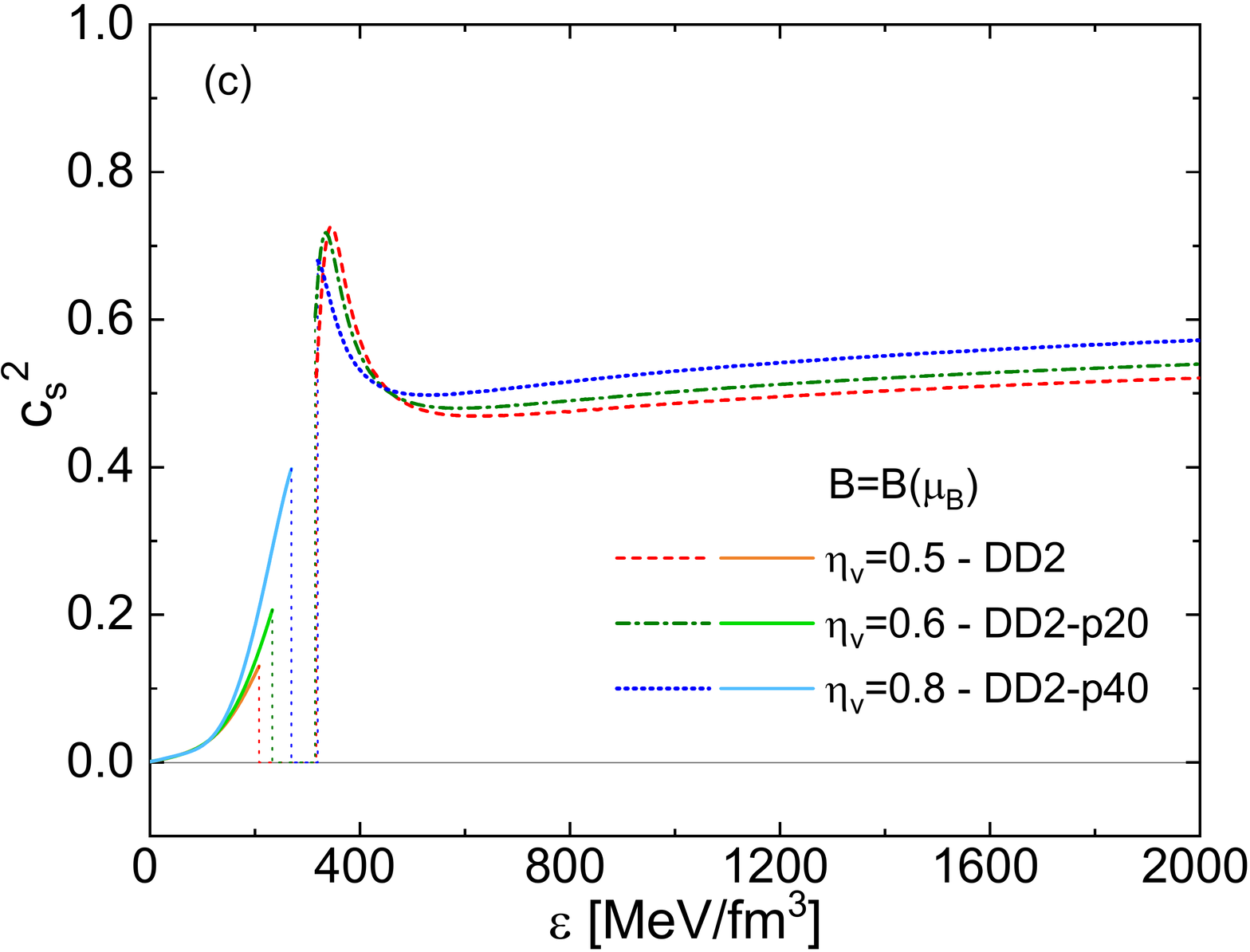}
    \includegraphics[width=0.49\textwidth]{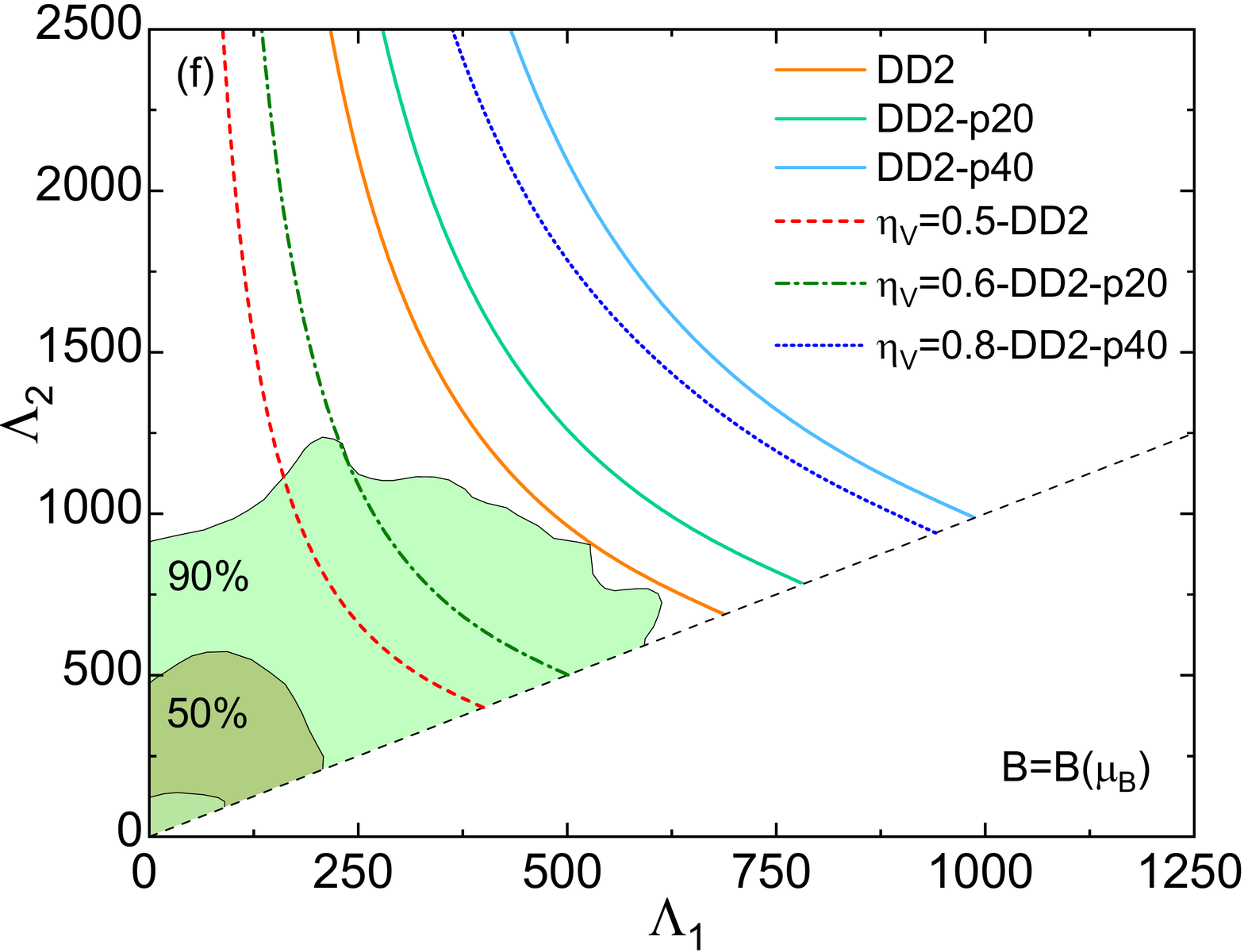}
    \caption{Same as Fig.~\ref{fig:all_B10}, but for the $\mu_B$-dependent bag pressure $B(\mu_B)$.
    }
    \label{fig:all_Bmu}
\end{figure}

\end{widetext}

This particular functional $\mu_B$ dependence of $B(\mu_B)$ permits to soften the QM EOS near the phase transition to avoid the masquerade problem \cite{Alford:2004pf}, and then, for higher $\mu_B$, $B(\mu_B) \rightarrow 0$ allowing for a stiffer EOS, necessary to reach the maximum mass constraint.

Besides, in order to accomplish the tidal deformability constraints, the hybrid EOS should be able to produce $M$-$R$ relations that fulfill the radius estimations for $M=1.4 M_{\odot}$ from the binary NS merger GW170817 \cite{LIGOScientific:2018cki}.

In Fig.~\ref{fig:all_Bmu} we summarize the results obtained with the proposed hybrid construction but considering now the $B(\mu_B)$ function instead of the constant $B$.
In Fig.~\ref{fig:all_Bmu}(a) we show $P$ vs $\mu_B$ for QM and hadronic EOS. 
For QM we display the corresponding EOS considering $\eta_D$ = 1.1 and three different values for $\eta_V$ = 0.5, 0.6, 0.8.
The hadronic EOS correspond to DD2, DD2-p20 and DD2-p40. It is observed that with the chosen parameter values the crossing of the hadronic and QM EOS occur at lower $P$ and $\mu_B$ values in comparison with the constant $B$ case.
In Fig.~\ref{fig:all_Bmu}(b), the hybrid EOS is included with NICER constraints, where a notorious change can be observed  with respect to the corresponding panel of Fig.~\ref{fig:all_B10}: now the energy gaps are bigger for lower values of $\eta_V$.
A consequence of the energy gap enhancement could be that the $M$-$R$ curves obtained in the QM phase have a bigger deviation from the corresponding hadronic lines around the phase transition. 
In Fig.~\ref{fig:all_Bmu}(c) the corresponding curves for the squared speed of sound are shown, where a peak is observed in the region of the transition from hadronic to QM.
A constant behavior at high densities, with $c_s^2$ between 0.5 and 0.6, can be seen as well.
In Fig.~\ref{fig:all_Bmu}(d) the corresponding mass-radius relations are displayed, showing both the hybrid solutions and the hadronic ones (lighter colored lines).
As in the previous section, different constraint regions from astrophysical observables are included. Earlier phase transition onset values compared with the ones with constant $B$ are obtained and, as in all the previous $M$-$R$ plots, the lower the vector coupling, the smaller the radius and the mass of the corresponding maximum mass stars. 
The tidal deformabilities $\Lambda(M)$ we obtained and the one corresponding to GW170817 are shown in Fig.~\ref{fig:all_Bmu}(e).
Finally, in Fig.~\ref{fig:all_Bmu}(f) the $\Lambda_1$-$\Lambda_2$ relations are shown and compared with the $50\%$ and $90\%$ probability regions from the inspiral gravitational wave measurement of GW170817.
As can be seen from the $B(\mu_B)$ results, it is possible to simultaneously fulfill the maximum mass and tidal deformability constraints.

\section{Summary and Conclusions}
In the present work we investigated the relationship between modern NS phenomenology and a microphysical approach formulated by a chiral quark model Lagrangian. 
We used a two-phase description of quark-nuclear matter to obtain a first-order deconfinement phase transition by a Maxwell construction between the DD2 model (with excluded volume and a crust of neutron stars) for nuclear matter and a nonlocal instantaneous NJL type model, including color superconductivity and local vector repulsive interactions for QM. 
Nonlocality was introduced in the quark currents through a Gaussian form factor in three-dimensional (3D) momentum space, except in the vector channel.
To obtain the input parameters of the QM model we first performed a fit to the momentum dependence of the quark mass function from LQCD in the Coulomb gauge to obtain a value for the effective range parameter $\Lambda_0$. 
Then, the values of the scalar coupling $G_S$ and the current quark mass $m_c$ were fixed by low-energy phenomenology. 
An advantage of the 3D-FF used in the present work is that it allows to perform Matsubara summations analytically. 

The first nontrivial result of this nonperturbative quark matter model is a coexistence phase of color superconductivity and chiral symmetry breaking that is due to strong coupling in the vector meson and diquark channels. 
The second remarkable finding obtained within the present QM model is the approximate constancy of the squared sound speed $c_s^2$ in the whole range of energy densities relevant for the quark matter interior of NS.
The approximately constant values for $c_s^2$ lie in the range $0.4-0.6$ for the considered quark model parameters. 
This confirms for the 3D-FF model the observation made earlier in Ref. \cite{Antic:2021zbn} for the 4D-FF model that a constant speed of sound model for quark matter with $c_s^2=0.45-0.54$ provides a perfect fit in the energy density range relevant for NS interiors. 
Our results show that we need to include a bag pressure in the QM EOS in order to fulfill modern observational constraints in a satisfactory way. 
Moreover, a $\mu_B$-dependent bag pressure is essential to simultaneously satisfy both radius constraints, for $R_{2.0}$ for high-mass NSs from the NICER observation of PSR J0740+6620 and $R_{1.4}$ for typical-mass NSs from the tidal deformability obtained from the inspiral gravitational wave measurement of GW170817. 
The bag pressure value close to the transition is in the range adopted by some works in the literature.

The hybrid description studied in the present work is the first step of testing the model before applying its finite temperature generalization to the simulation of supernova explosions and NS mergers  \cite{Fischer:2017lag}.
We expect to report on this issue in the near future.

\subsection*{Acknowledgements}
A.G.G., G.A.C., and J.P.C. would like to acknowledge CONICET and UNLP for financial support under Grants No. PIP17-700, No. G157, and No. X824. 
J.P.C. and G.A.C. acknowledge support from CONICET under grant P-UE 2016, ``B\'usqueda de Nueva F\'isica''.
The work of D.B. was supported by the Polish National Science Centre (NCN) under grant No. 2019/33/B/ST9/03059, by the Russian Fund for Basic Research (RFBR) under grant No. 18-02-40137 and by the Russian Federal Program ``Priority-2030''.
This work is part of a project that has received funding from the European Union’s Horizon 2020 research and innovation program under the grant agreement STRONG – 2020, No. 824093.

\begin{appendix}
\section{Details of the nonlocal model for quark matter}
\label{sect:details_QM}

In this Appendix we show some explicit expressions
corresponding to the nonlocal chiral quark model considered in
Sec.~\ref{sect:3DFFqm}.

From Eqs. (\ref{gapeq}) the gap equations for the mean fields $\bar{\sigma}$ and $\bar{\Delta}$ together with the constraint equation for the mean field $\bar\omega$ read
\begin{eqnarray}
\bar\sigma &=& 2 G_s \int \frac{d^3\vec{p}}{(2 \pi)^3} 
\frac{g(\vec{p})\ M(\vec{p})}{E}  \\ 
& &\times \sum_{\kappa} 
\left\{1-n_F(\frac{\bar{E}^\kappa_b + \delta\tilde\mu_b}{T})
- n_F(\frac{\bar{E}^\kappa_b - \delta\tilde\mu_b}{T}) \right.  \nonumber\\
&+& \left.\frac{2 \bar{E}^\kappa_r}{\epsilon^\kappa_r}\left[ 1 - n_F(\frac{\epsilon^\kappa_r + \delta\tilde\mu_r}{T})
- n_F(\frac{\epsilon^\kappa_r - \delta\tilde\mu_r}{T}) \right] \right\},\nonumber 
\label{gap_sig}
\end{eqnarray}

\begin{eqnarray}
\bar\Delta &=&  2 G_D \bar\Delta \int \frac{d^3\vec{p}}{(2 \pi)^3} 
g^2(\vec{p})   \\
&&\times  \sum_{\kappa=\pm} 
\left\{ \frac{2}{\epsilon^\kappa_r}\left[ 1 - n_F(\frac{\epsilon^\kappa_r + \delta\tilde\mu_r}{T})
- n_F(\frac{\epsilon^\kappa_r - \delta\tilde\mu_r}{T}) \right] \right\},\nonumber
\label{gap_del}
\end{eqnarray}
and
\begin{eqnarray}
\bar\omega &=& 2 G_V \int \frac{d^3\vec{p}}{(2 \pi)^3}  \\ 
&&\times \sum_{\kappa=\pm} \kappa
\left\{ 1 - n_F(\frac{\bar{E}^\kappa_b + \delta\tilde\mu_b}{T})
- n_F(\frac{\bar{E}^\kappa_b - \delta\tilde\mu_b}{T}) \right. \nonumber\\
&+& \left.\frac{2 \; \bar{E}^\kappa_r}{\epsilon^\kappa_r}\left[ 1 - n_F(\frac{\epsilon^\kappa_r + \delta\tilde\mu_r}{T})
- n_F(\frac{\epsilon^\kappa_r - \delta\tilde\mu_r}{T}) \right] \right\}, \nonumber
\label{gap_ovec}
\end{eqnarray}
where $n_F(x)=(1+\exp(x))^{-1}$ is the Fermi distribution function.
The solutions for Eqs. (\ref{gap_sig})-(\ref{gap_ovec}) in the vacuum, at $T=\mu=0$, are denoted with the subscript $0$ as 
$\bar \sigma_0$, $\bar \Delta_0$ and $\bar \omega_0$, respectively.
In the vacuum $\bar \Delta_0$ and $\bar \omega_0$ are zero, so the vacuum thermodynamic potential reads
\begin{eqnarray}
\Omega^{MFA}_0 = \frac{ \bar
\sigma^2_0 }{2 G_S} - 12 \int \frac{d^3 \vec{p}}{(2\pi)^3} \; E_0/2. 
\label{mfavac}
\end{eqnarray}
Notice that in the above expression $E_0^2 = \vec{p}~^2 + M_0^2(\vec{p})$ where $M_0(\vec{p}) = m_c + g(\vec{p}) \bar\sigma_0$.
The integral in Eq.~(\ref{mfaqmtp}) turns out to be ultraviolet divergent because of the zero-point energy terms. Since this is exactly the divergence of Eq.~(\ref{mfavac}), a successful regularization scheme consists just in the vacuum subtraction
\begin{equation}
\Omega^{MFA}_{reg} = \Omega^{MFA} - \Omega^{MFA}_0 \,.
\label{eq:omreg}
\end{equation}
Finally, the quarks and lepton number densities, $n_{fc}$ and $n_l$ respectively,
are given by
\begin{eqnarray}
n_{fc}  =  {-\; \frac{ \partial \Omega^{MFA}}{\partial \mu_{fc}} \;
}
\; \ , \;\;\;
n_l = - \frac{ \partial \Omega_{lep}}{\partial \mu_{e}} \ ,
\label{densities_2}
\end{eqnarray}
where $f=u,d$, $c=r,g,b$, and $l=e,\mu$. Thus,
\begin{eqnarray}
n_{fr} &=& n_{fg} = 
 - \int \frac{d^3 \vec{p}}{(2\pi)^3} \;  \\
& &\times \sum_{\kappa=\pm} \left\{\left[  n_F(\frac{\epsilon_r^\kappa + \delta \tilde\mu_r}{T}) -  n_F(\frac{\epsilon_r^\kappa - \delta \tilde\mu_r}{T})\right] (\delta_{uf} - \delta_{df}) \right.\nonumber \\
&-& \left. \kappa \;
\frac{\bar{E}_r^\kappa}{\epsilon_r^\kappa} \left[ 1 -  n_F(\frac{\epsilon_r^\kappa +  \delta\tilde\mu_r}{T}) - n_F(\frac{\epsilon_r^\kappa -  \delta\tilde\mu_r}{T})
 \right]\right\}, \nonumber
\label{rhofr}
\end{eqnarray}

\begin{eqnarray}
n_{fb} &=& 
- \int \frac{d^3 \vec{p}}{(2\pi)^3} \;  \\
& &\times \sum_{\kappa=\pm} \left\{\left[  n_F(\frac{\bar{E}_b^\kappa + \delta \tilde\mu_b}{T}) -  n_F(\frac{\bar{E}_b^\kappa - \delta \tilde\mu_b}{T})\right] (\delta_{uf} - \delta_{df}) \right.\nonumber \\
&-& \left. \kappa \;
 \left[ 1 -  n_F(\frac{\bar{E}_b^\kappa +  \delta\tilde\mu_b}{T}) - n_F(\frac{\bar{E}_b^\kappa -  \delta\tilde\mu_b}{T})
 \right]\right\}, \nonumber 
\label{rhofb}
\end{eqnarray}
and
\begin{eqnarray}
n_l &=& - 2  \;  \sum_{l = e, \mu}
\int \frac{d^3 \vec{p}}{(2\pi)^3} \;  \\
& &\times \left[ n_F(\frac{\epsilon_l +  \; \mu_e}{T}) - n_F(\frac{\epsilon_l -  \; \mu_e}{T})\right]. \nonumber
\label{rholep}
\end{eqnarray}   
Note that the fermion density has only the contribution of the first term of the regularized thermodynamic potential (\ref{eq:omreg}), since the second one has no dependence on the chemical potentials.

\vspace{0.5cm} 
In what follows we present some analytic expressions in the zero temperature limit, where for notation convenience we have omitted the subscript $T=0$.

The corresponding gap equations and constraint equation are given by
\begin{eqnarray}
\bar\sigma &=& 2 G_S \int \frac{d^3 \vec{p}}{(2\pi)^3} \; 
\frac{g(\vec{p})\ M(\vec{p})}{E}  \\ 
& &\times \sum_{\kappa,s=\pm} 
\left\{ \frac{\bar{E}_r^{\kappa}}{\epsilon_r^{\kappa}} 
\frac{\epsilon_r^{\kappa} +s\ \delta\tilde\mu_r}{|\epsilon_r^{\kappa} +s\ \delta\tilde\mu_r|} 
\right. 
+ \left. \frac{1}{2}\frac{\bar{E}_b^{\kappa} +s\ \delta\tilde\mu_b}{|\bar{E}_b^{\kappa} +s\ \delta\tilde\mu_b|}  \right\}, \nonumber
\end{eqnarray}

\begin{eqnarray}
\bar\Delta &=& 2 G_D \bar{\Delta}\int \frac{d^3 \vec{p}}{(2\pi)^3} \; g^2(\vec{p})
\sum_{\kappa,s=\pm} 
\frac{1}{\epsilon_r^{\kappa}} \frac{\epsilon_r^{\kappa} + s\ \delta\tilde\mu_r}{|\epsilon_r^{\kappa} + s\ \delta\tilde\mu_r|}, \nonumber\\
\end{eqnarray}

\begin{eqnarray}
\bar\omega &=& 2 G_V \int \frac{d^3 \vec{p}}{(2\pi)^3} \; 
 \\ 
& &\times \sum_{\kappa,s=\pm} \kappa
\left\{ 
\frac{\bar{E}_r^{\kappa}}{\epsilon_r^{\kappa}} 
\frac{\epsilon_r^{\kappa} +s\ \delta\tilde\mu_r}{|\epsilon_r^{\kappa} +s\ \delta\tilde\mu_r|} 
+ 
\frac{1}{2}\frac{\bar{E}_b^{\kappa} +s\ \delta\tilde\mu_b}{|\bar{E}_b^{\kappa} +s\ \delta\tilde\mu_b|} \right\}. \nonumber
\end{eqnarray}
The number densities read
\begin{eqnarray}
n_{fr} &=& n_{fg} = \int \frac{d^3 \vec{p}}{(2\pi)^3} \times \nonumber \\
&&\times \sum_{\kappa=\pm} \left\{
 \frac{\epsilon_r^{\kappa} + \delta\tilde\mu_r}{|\epsilon_r^{\kappa} + \delta\tilde\mu_r|} 
\frac{1}{2}\left[\kappa \frac{\bar{E}_r^{\kappa}}{\epsilon_r^{\kappa}} + (\delta_{uf} - \delta_{df}) \right] 
\right. \nonumber\\
&+& \left.
\frac{\epsilon_r^{\kappa} - \delta\tilde\mu_r}{|\epsilon_r^{\kappa} - \delta\tilde\mu_r|} 
\frac{1}{2}\left[\kappa \frac{\bar{E}_r^{\kappa}}{\epsilon_r^{\kappa}} - (\delta_{uf} - \delta_{df}) \right] 
\right\},
\end{eqnarray}

\begin{eqnarray}
n_{fb} &=& \int \frac{d^3 \vec{p}}{(2\pi)^3} \; 
\sum_{\kappa=\pm} \left\{
 \frac{\bar{E}_b^{\kappa} + \delta\tilde\mu_b}{|\bar{E}_b^{\kappa} + \delta\tilde\mu_b|} 
\frac{1}{2}\left[\kappa + (\delta_{uf} - \delta_{df}) \right] \right. \nonumber\\
&& + \left.
\frac{\bar{E}_b^{\kappa} - \delta\tilde\mu_b}{|\bar{E}_b^{\kappa} - \delta\tilde\mu_b|} \frac{1}{2}\left[\kappa - (\delta_{uf} - \delta_{df}) \right]
\right\},
\end{eqnarray}
where for leptons
\begin{eqnarray}
n_l \ = \ \frac{1}{3\pi^2} \;\,
(\mu_l^2 - m_l^2)^{3/2}.
\end{eqnarray}
Finally, for completeness, the chiral quark condensate is defined as
\begin{equation}
\langle \bar q q \rangle  =  \frac{ \partial \Omega^{MFA}_{reg}}{\partial m_c} = -\int \frac{d^3 \vec{p}}{(2\pi)^3} \ \zeta(\vec{p}),  \;
\end{equation}

where
\begin{eqnarray}
\zeta(\vec{p}) &=& \sum_{\kappa,s=\pm} \left\{ 
2\left[(\mathrm{sgn}(\epsilon^{\kappa}_r +s\ \delta\tilde\mu_r)\frac{M(\vec{p})}{E} \frac{\bar E^{\kappa}_r}{\epsilon^{\kappa}_r}- \frac{M_0(\vec{p})}{E_0} \right] \right.\nonumber\\
&+& 
\left.  \left[(\mathrm{sgn}(\bar E^{\kappa}_b + s\ \delta\tilde\mu_b)\frac{M(\vec{p})}{E} - \frac{M_0(\vec{p})}{E_0} \right]  \right \}.
\end{eqnarray}

\section{Calculation of gravitational mass, radius, and tidal deformability of spherical compact stars}
\label{sect:TOV_tidals}

To compute the internal energy density distribution of compact stars and thus derive the mass-radius relation we utilize the TOV equations for a static and spherical star
in the framework of general relativity:
\begin{align}
    \frac{dP(r)}{dr} &= \frac{G(\epsilon(r))+P(r))(M(r)+4\pi r^3 P(r))}{r(r-2GM(r))},\\
    \frac{dM(r)}{dr} &= 4\pi r^2 \epsilon(r)\,,
    \label{eq:TOV}
\end{align}
with $P(r=R)=0$ and $P(r=0)=P_c$ as boundary conditions for a star with mass $M$ and radius $R$.

Here we want to briefly describe how to compute the tidal deformability (TD) of a compact star, based on the results of Refs.  ~\cite{Hinderer:2007mb,Damour:2009vw,Binnington:2009bb,Hinderer:2009ca,Yagi:2013awa}.
To determine the dimensionless tidal deformability  parameter $\Lambda=\lambda/M^{5}$ that can be computed for small tidal deformabilities in a perturbative way,
here $\lambda$ is the stellar TD and $M$ is the stellar gravitational mass\,\cite{Alvarez-Castillo:2018pve}.
In addition, $\lambda$ is related to the so called Love number:
\begin{eqnarray}
  k_2 = \frac{3}{2} \lambda R^{-5}.
  \label{eq:k2}
\end{eqnarray}
The TD can be thought of as a modification of the space-time metric by a linear $l=2$ perturbation of a spherical symmetric star,
\begin{eqnarray}
\begin{split}
    ds^2 =& - e^{2\Phi\left(r\right)} \left[1 + H\left(r\right) Y_{20}(\theta,\varphi)\right]dt^2 \\
  &+ e^{2\Lambda\left(r\right)} \left[1 - H\left(r\right) Y_{20}\left(\theta,\varphi\right)\right]dr^2 \\
 &+ r^2 \left[1-K\left(r\right) Y_{20}\left(\theta,\varphi\right)\right] \left( d\theta^2+ \sin^2\theta
 d\varphi^2 \right),
\end{split}
\end{eqnarray}
where $K'(r)=H'(r)+2 H(r) \Phi'(r)$, primes denoting derivatives with respect to $r$.

The functions $H(r)$, $\beta(r) = dH/dr$ obey
 \begin{align}
  \frac{dH}{dr}=& \beta,\\
  \begin{split}
       \frac{d\beta}{dr}=&2 \left(1 - 2\frac{m(r)}{r}\right)^{-1} \\
  & \times H\left\{-2\pi
    \left[5\varepsilon(r)+9 P(r)+f(\varepsilon(r)+P(r))\right]\phantom{\frac{3}{r^2}} \right. \\
 & \left.+\frac{3}{r^2}+2\left(1 - 2\frac{m(r)}{r}\right)^{-1}
    \left(\frac{m(r)}{r^2}+4\pi r P(r)\right)^2\right\}\\
 &+\frac{2\beta}{r}\left(1 -
   2\frac{m(r)}{r}\right)^{-1}\\
 & \times \left\{-1+\frac{m(r)}{r}+2\pi r^2
 (\varepsilon(r)-P(r))\right\}~,
  \end{split}
 \end{align}
 where $f =d\epsilon/dp$ is the equation of state.
  The above equations and the TOV equations have to be solved simultaneously.
 The system is to be integrated outward starting near the center using the expansions $H(r)=a_0 r^2$ and $\beta(r)=2a_0r$ as $r \to 0$, where  $a_0$ is a constant that determines how much the star is deformed and turns out to be arbitrary since it cancels in the expression for the Love number.
 With the definition of
\begin{align}
 y = \frac{ R\, \beta(R)} {H(R)},
\end{align}
 the $l=2$ Love number is found as
\begin{align}
\begin{split}
     k_2 =& \frac{8C^5}{5}\left(1-2C\right)^2\left[2+2C\left(y-1\right)-y\right] \\ 
     &\times \Big\{2C[6-2y+3C(5y-8)] \\
     & + 4C^3\left[13-11y+C\left(3y-2\right)+2C^2(1+y)\right]\\
     & +3\left(1-C\right)^2\left[1-y+2C\left(y-1\right)\right]\ln{1-2C}\Big\}^{-1},
\end{split}
\end{align}
where $C=M/R$ is the compactness of the star.

Finally, note that the $\Lambda_1$ and $\Lambda_2$ parameters for the two components of the NS merger are obtained from Eq.~(\ref{eq:k2}) with the corresponding $M$-$R$ values, 
\begin{align} 
   \Lambda_{1,2}= \frac{2}{3} k_2  \left(\frac{R_{1,2}}{M_{1,2}}\right)^{5}.
\end{align}
\end{appendix}
\bibliographystyle{apsrev4-2}
\bibliography{3DFF}

\end{document}